# Estimating the impact of structural directionality: How reliable are undirected connectomes?

Penelope Kale[1, 2], Andrew Zalesky[3], Leonardo L. Gollo[1, 2]

1. QIMR Berghofer Medical Research Institute, Australia
2. The University of Queensland, Australia
3. Melbourne Neuropsychiatry Centre and Department of Biomedical Engineering, The University of Melbourne, Australia

**Abstract**

Directionality is a fundamental feature of network connections. Most structural brain networks are intrinsically directed because of the nature of chemical synapses, which comprise most neuronal connections. Due to limitations of non-invasive imaging techniques, the directionality of connections between structurally connected regions of the human brain cannot be confirmed. Hence, connections are represented as undirected, and it is still unknown how this lack of directionality affects brain network topology. Using six directed brain networks from different species and parcellations (cat, mouse, C. elegans, and three macaque networks), we estimate the inaccuracies in network measures (degree, betweenness, clustering coefficient, path length, global efficiency, participation index, and small worldness) associated with the removal of the directionality of connections. We employ three different methods to render directed brain networks undirected: (i) remove uni-directional connections, (ii) add reciprocal connections, and (iii) combine equal numbers of removed and added uni-directional connections. We quantify the extent of inaccuracy in network measures introduced through neglecting connection directionality for individual nodes and across the network. We find that the coarse division between core and peripheral nodes remains accurate for undirected networks. However, hub nodes differ considerably when directionality is neglected. Comparing the different methods to generate undirected networks from directed ones, we generally find that the addition of reciprocal connections (*false positives*) causes larger errors in graph-theoretic measures than the removal of the same number of directed connections (*false negatives*). These findings suggest that directionality plays an essential role in shaping brain networks and highlight some limitations of undirected connectomes.

**Introduction**

Connectomes provide a comprehensive network description of structural brain connectivity (Sporns et al., 2005). Large-scale connectomes mapped in humans are typically represented and analyzed as undirected networks, due to the inability of non-invasive connectome mapping techniques to resolve the directionality (afferent or efferent) of white matter fibers. Reducing an inherently directed network such as the connectome to an undirected network is a simplification that may introduce inaccuracies in graph-theoretic analyses. For example, the flow of action potentials along an axon is mostly only ever in one direction, and thus analyses of information flow are critically dependent on connection directionality. This study aims to systematically and comprehensively characterize the impact of representing and analyzing connectomes as undirected networks.

At the neuronal level, the connections between nodes (neurons) are given by synapses, and the great majority of them are chemical, which have distinctive pre- and post-synaptic terminals determining the direction of neurotransmitter flux (Kandel et al., 2000). This structural feature of chemical synapses emphasizes the importance of directionality for the connections, and therefore for the whole network. Invasive techniques to map connectomes such as tract tracing (Scannell et al., 1999, Kötter, 2004, Sporns et al., 2007, Oh et al., 2014) or electron microscopy (White et al., 1986, Achacoso and Yamamoto, 1992) can detect the directionality of the connections. Conversely, human connectomes are currently mapped with non-invasive tractography methods performed on diffusion-weighted magnetic resonance imaging data (Assaf and Basser, 2005, Hagmann et al., 2008, Tournier et al., 2012). While methods for improving the quality of diffusion-based connectomes have advanced in recent years, and numerous tractography algorithms have been developed to reconstruct axonal fiber bundles, they cannot provide any information about the directionality of the connections. Therefore, analyses of the human connectome, as well as modeling studies that use the human connectivity matrix, are compromised by the lack of information regarding directionality, which is one of the most fundamental features of complex networks.

In the absence of directionality, networks are considered undirected and therefore the connections only represent the existence of a relationship between nodes. This is the case for scientific co-authorship networks (Newman, 2004), film actor networks (Watts and Strogatz, 1998), and functional networks defined by symmetric functions such as the Pearson correlation (Biswal et al., 1995) or the phase locking value (Aydore et al., 2013). Among others, studies of tractography-derived human brain networks have revealed a variety of important features such as hub regions (van den Heuvel and Sporns, 2013), modularity and clustering (Sporns, 2011, Sporns and Betzel, 2016), small worldness (Bassett and Bullmore, 2006, Medaglia and Bassett, 2017), core-periphery structure (Hagmann et al., 2008) and the existence of a rich club (van den Heuvel and Sporns, 2011). These topological properties are not specific to the human brain. Comparisons across many species have recapitulated these features (Harriger et al., 2012, Towlson et al., 2013,

Betzel and Bassett, 2016, van den Heuvel et al., 2016). However, the topological characteristics of connectomes, as well as many other graph-theoretic measures, are affected by the directionality of connections (Rubinov and Sporns, 2010).

When directionality cannot be identified, undirected representations of connectomes are incomplete. Undirected networks inform the presence of a relationship between two brain regions. But these networks lack information about the asymmetry of this relationship. For example, if a directed network is represented as an undirected network, uni-directional connections are either present, which can be interpreted as a spurious addition of a reciprocal connection (*false positives*), or overlooked (*false negatives*). More specifically, if a uni-directional connection exists from node $u$ to $v$, but not from $v$ to $u$, then the undirected representation of this connection is either: (i) an undirected connection between $u$ and $v$, which can be construed as admitting a false positive from node $v$ to $u$; or, (ii) absence of an undirected connection between $u$ and $v$, which can be construed as a false negative from node $u$ to $v$. In either case, a potential error (false positive or false negative) is introduced to the undirected network.

Beyond the effect of directionality, connectomes also contain errors in the balance between overlooked and spurious connections owing to imprecisions in currently available mapping techniques (Calabrese et al., 2015, Donahue et al., 2016). Although both error types impact the network topology, spurious (false positive) connections introduce inaccuracies in a few graph-theoretic measures (network clustering, efficiency and modularity) in different connectomes that are at least twice as large as those found with the same number of overlooked (false negative) connections (Zalesky et al., 2016). This finding indicates that the importance of specificity is much greater than sensitivity for general connectivity in which false positives could be any absent connection and false negatives, any present connection. However, the impact of representing a directed connection as undirected, which, for practical purposes, is typically indistinguishable from a bidirectional connection, is currently unknown. Therefore, when directed networks are mapped with techniques that cannot infer directionality, it is important to establish what undirected representation is the most detrimental with respect to directionality: admitting spurious reciprocal connections (false positives) or overlooking uni-directional connections (false negatives).

Moreover, the effect of directionality on the identification of network hubs may also be important as hubs play an important role for normal brain function (van den Heuvel et al., 2012, Mišić et al., 2015) as well as in neuropsychiatric disorders (Bassett et al., 2008, Crossley et al., 2014, Fornito et al., 2015). But how are these highly connected regions affected by directionality? Does the classification of nodes into hubs still hold if directionality is taken into account? Furthermore, to what extent do graph-theoretic measures at the node level remain valid? The characterization of the human brain as an undirected network is often overlooked and requires investigation.

The aim of this study is to understand the limitations of analyzing inherently directed connectomes as undirected networks. Beginning with directed connectomes of the macaque, cat, mouse, and Caenorhabditis elegans (C. elegans), we study how seven graph-theoretic measures are affected as we progressively modify uni-directional connections, either deleting them or making them undirected. More specifically, we consider three schemes to progressively eliminate directionality information: removing uni-directional connections (creating false negatives), adding reciprocal connections to existing uni-directional connections (creating false positives), and removing one uni-directional connection for each reciprocal connection added, thus preserving the density and mean degree of the original network. We show how essential network features, such as the identification and classification of hubs, are affected by perturbations in directionality. Moreover, we quantify how graph-theoretic measures are affected at both the node and network level and determine whether false positive or false negative uni-directional connections are more detrimental to the characterization of graph-theoretic measures.

**Materials and Methods**

*Connectivity Data*

Following a comparative connectomics approach (van den Heuvel et al., 2016), we analyzed structural connectivity data from several species and various parcellations including three macaque connectomes, a cat and mouse connectome, and a C. elegans nervous system connectome (Fig. 1). Each network possesses a different number of nodes, proportion of uni-directional connections, modularity, and network density (see Supplementary Table 1). Crucially, these networks include information on the directionality of connections (all networks are directed) obtained through invasive techniques that have different proportions of connection reciprocity (Garlaschelli and Loffredo, 2004). Among the meso- and macro-scale connectomes, nodes represent cortical regions and the directed connections represent axons or white matter fibers linking these regions via chemical synapses. In the case of the micro-scale C. elegans connectome, nodes represent neurons, the directed connections represent chemical synapses, and the electrical synapses (or gap junctions) are bidirectional connections.

To accommodate the analysis of such a wide range of directed connectomes, the strength of connections was disregarded (for the cat and mouse connectomes) to make each network binary. This procedure allowed us to characterize all connectomes using the same methods for binary and directed networks as a first step to understand the role of directionality in structural brain networks. Other high-quality weighted connectomes can be used in future studies (Bezgin et al., 2012, Markov et al., 2012, Shih et al., 2015, Ypma and Bullmore, 2016, Gămănuţ et al., 2017). As recently reported, the combination of both directionality and weight can be crucial to uncover relationships between structural connectivity and univariate brain dynamics (Sethi et al., 2017).

**Macaque networks:** The first macaque network, used in a study by Honey et al. (2007), (with number of nodes N = 47 and connections E = 505, Fig. 1A) follows the parcellation scheme of Felleman and Van Essen (1991) including the visual and sensorimotor cortex, and motor cortical regions. Relevant data was collated in the CoCoMac database (Modha and Singh, 2010) following the procedures of Kötter (2004) and Stephan et al. (2001), and translated to the brain map using coordinate independent mapping (Stephan et al., 2000, Kötter and Wanke, 2005).

The second macaque connectome (N = 71 and E = 746, Fig. 1B) was derived from a whole cortex model generated by Young (1993) with regions of the hippocampus and amygdala eliminated. The parcellation was based mostly on the scheme by Felleman and Van Essen (1991), except for the fields of the superior temporal cortex (Yeterian and Pandya, 1985). Yeterian and Pandya (1985) utilized an autoradiographic technique (radioactively labeled amino acids) to establish the existence and trajectory of fibers.

The final macaque connectome (N = 242 and E = 4090, Fig. 1C) was generated by Harriger et al. (2012). This network comprises anatomical data from over 400 tract tracing studies collated in the CoCoMac database (Modha and Singh, 2010) following the procedures of Kötter (2004) and Stephan et al. (2001), focusing on the right hemisphere with all subcortical regions removed as well as regions without at least one incoming and one outgoing connection.

The data collated for the CoCoMac database used a range of tracer substances (with anterograde, retrograde or bidirectional transport properties) and methods (as discussed in Stephan et al. (2001)). Each contributing study must discern a source and target for the connection. If the reciprocal direction had not been tested for, the connection was assumed to be uni-directional. Some connections have been confirmed to be uni-directional, for example, the connection from V2 to FST, see Boussaoud et al. (1990). Regarding macaque connectomes Felleman and Van Essen (1991) have also suggested that the reciprocity of connections may vary between individuals.

**Cat network:** The cat matrix is a connectome reconstructed by Scannell et al. (1999) and curated from a database of thalamo-cortico-cortical connections from a large number of published studies in the adult cat. The parcellation was based on a previous scheme by Reinoso-Suarez (1984) and adapted by Scannell et al. (1995). Areas ALG, SSF, SVA, DP, Amyg and 5m were discarded (and some regions grouped) to create a weighted network (N = 52 and E = 818, Fig. 1D). This connectome was generated from the available data across numerous studies. It is noted that each study used a different type of anterograde and/or retrograde tracer, methodology and parcellations. Some connections lacked data on the existence of a reciprocal direction between brain regions (these were left as uni-directional), and all connections between the cortex and thalamus were assumed to be reciprocal.

**Mouse network:** We obtained the mouse connectome (N = 213 and E = 2105, Fig. 1E) from the Allen Mouse Brain Connectivity Atlas generated by Oh et al. (2014). The major advantage of this connectome is that the connectivity data, obtained at a cellular level (axons and synaptic terminals), is generated for the whole mouse brain. Therefore, all 469 individual experiments use the same anterograde tracer and consistent techniques. Each brain is applied to a 3D template, which itself is averaged across 1231 brain specimens, and the regions are matched against the Allen reference atlas (Oh et al., 2014). We thresholded this dense and weighted network using the disparity filter (Serrano et al., 2009) maintaining only connections with a p-value smaller than 0.05. Thresholding was performed such that the resulting network was binary.

**C. elegans network:** The C. elegans nervous system matrix (N = 279 and E = 1943, Fig. 1F) was collated by Varshney et al. (2011), and includes data mapped by White et al. (1986) using electron microscopy, in addition to various other sources (White et al., 1976, Durbin, 1987, Hall and Russell, 1991). This microscale connectome is comprised of a directed chemical synapse network and an undirected gap junction network. Although gap junctions may possess directionality, this has not yet been demonstrated in C. elegans. For the purpose of analysis, the connections from the gap junction network were treated as bidirectional connections.

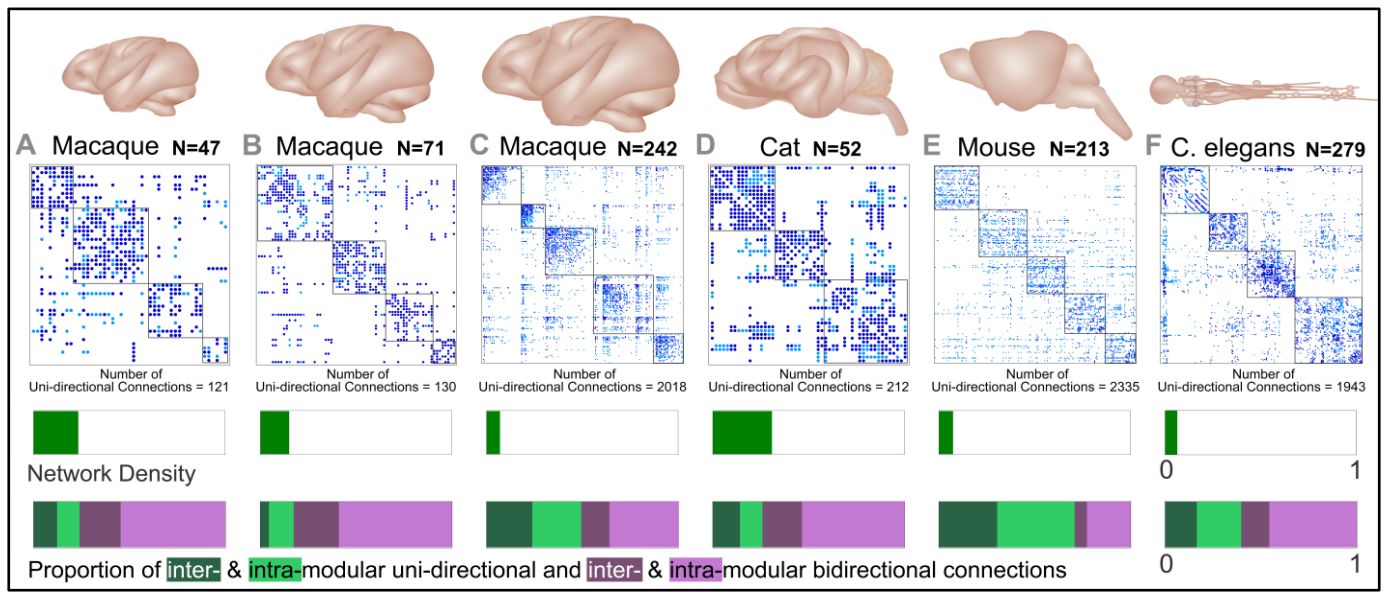

Figure 1: **The six connectomes analyzed in this study.** Brain and connectome for three different parcellations of the macaque cortex (A) nodes N=47 (Honey et al., 2007), (B) N=71 (Young, 1993), and (C) N=242 (Harriger et al., 2012), as well as three additional species including a (D) cat (Scannell et al., 1999), (E) mouse (Oh et al., 2014), and (F) C. elegans (White et al., 1986, Varshney et al., 2011). The connectomes represent connectivity matrices with rows and columns denoting brain regions (or nodes), and the elements within the matrices denoting the presence (filled) or absence (blank) of a connection between two regions. Uni-directional connections are highlighted in light blue (with the number of uni-directional connections stated below each connectome) and the nodal regions are arranged into modular communities. The bars below each connectome display the density of each network (A= 0.234, B= 0.15, C= 0.07, D= 0.308, E= 0.073, F=0.063) and the proportion of uni-directional and bidirectional connections. The latter is segmented to display the proportion of uni-directional connections between modules (dark green: A= 0.123, B= 0.046, C= 0.238, D= 0.142, E= 0.304, F= 0.165) and within modules (light green: A= 0.117, B= 0.129, C= 0.255, D= 0.117, E= 0.404, F= 0.232) separately, as well as proportion of bidirectional connections between modules (dark purple: A= 0.214, B= 0.236, C= 0.147, D= 0.21, E= 0.064, F= 0.147) and within modules (light purple: A= 0.547, B= 0.59, C= 0.359, D= 0.536, E= 0.229, F= 0.457).

*Perturbed Networks*

To investigate the effects of directionality on the characteristics of the brain, each empirical connectome was altered by progressively removing connection directionality information, generating a spectrum of perturbed networks. This spectrum comprised the empirical connectome at one end, and a fully undirected representation of the connectome at the opposite end. For this purpose, the empirical networks were considered to be approximately the ground-truth connectomes for a given parcellation. Figure 2 illustrates the three different approaches used to generate perturbed networks for the macaque (N=47) connectome. The empirical connectome is shown in Fig. 2A, and the uni-directional connections of this network are shown in Fig. 2B. Perturbed networks (Fig. 2C-E) were generated by altering the directionality or presence of the uni-directional connections. In this example, we only show the extreme case in which all information about connection directionality is removed, yielding a fully undirected perturbed network.

For further analyses we present three schemes that were developed to progressively eliminate connection directionality information from the empirical connectomes, yielding perturbed networks that increasingly resembled undirected networks.

**False negative perturbed networks:** The first perturbed network was generated by removing a fixed number of randomly chosen uni-directional connections, leading to a connectome with false negative uni-directional connections (FN network, Fig. 2C). The perturbed network was undirected in the extreme case when all uni-directional connections were removed. This perturbation assumes that uni-directional connections are weaker in strength (weight) relative to their bidirectional counterparts, and thus uni-directional connections are most vulnerable to elimination with weight-based thresholding procedures (Rubinov and Sporns, 2010). Such thresholding is commonly used to eliminate weak connections obtained with tractography, which are often attributed to noise or error (Maier-Hein et al., 2017). As an example, the majority of the weighted mouse connectome is comprised of uni-directional connections (57%), and they are also weaker than bidirectional connections. The mean of the strength of uni-directional connections is 0.066, whereas the mean strength of bidirectional connections is 0.165, which is significantly weaker ($P<10^{-45}$, Welch's t-test).

**False positive perturbed networks:** If the weight of a uni-directional connection exceeds the weight-based threshold, the connection will be represented in the perturbed network as an undirected connection (i.e. a uni-directional connection from node *u* to *v* becomes an undirected connection between nodes *u* and *v*). In this case, the undirected connection is treated as a bidirectional connection, and thus construed as a false positive. To model this case, we generated perturbed networks by adding reciprocal connections to a fixed number of randomly chosen existing uni-directional connections, leading to a perturbed network with false positive reciprocal connections (FP network, Fig. 2D). In the extreme case when all reciprocal connections were added, the perturbed network effectively became an undirected network.

**Density-preserving perturbed networks:** Finally, to preserve basic properties of the empirical connectome, an additional perturbed connectome termed the density-preserving network was generated (DP network, Fig. 2E). In this perturbed connectome, for each reciprocal connection added to a uni-directional connection, another uni-directional connection is removed (at randomly selected locations). The DP network has an equal number of false negative and positive connections and also preserves the mean degree of the empirical connectome, but not the degree of each node.

To generate undirected perturbed networks, we progressively applied one of the above three schemes to randomly chosen uni-directional connections in the empirical connectomes until a desired proportion of connections were changed. We generated perturbed networks in which 5%, 10%, 20% and 100% of directed connections were altered (eliminated or the reciprocal connection added). This process was repeated for multiple trials to generate an ensemble of perturbed networks. Ensemble averages for all graph-theoretic measures were then computed. Each perturbed network was associated with a rewiring scheme (FN, FP, and DP) and a proportion of changed connections. Supplementary Table 2 provides the details of the proportion of uni-directional connections altered in the perturbed networks and other relevant parameters used for each analysis.

The perturbed networks can comprise isolated nodes that are not connected to any other nodes (see Supplementary Fig. 1). Isolated nodes are more likely to occur in the FN perturbed networks, potentially having a greater impact on graph-theoretic measures as more connections are changed. Therefore, in cases where only a subset of uni-directional connections are modified (<100%), the trials that cause nodes to become disconnected are rejected.

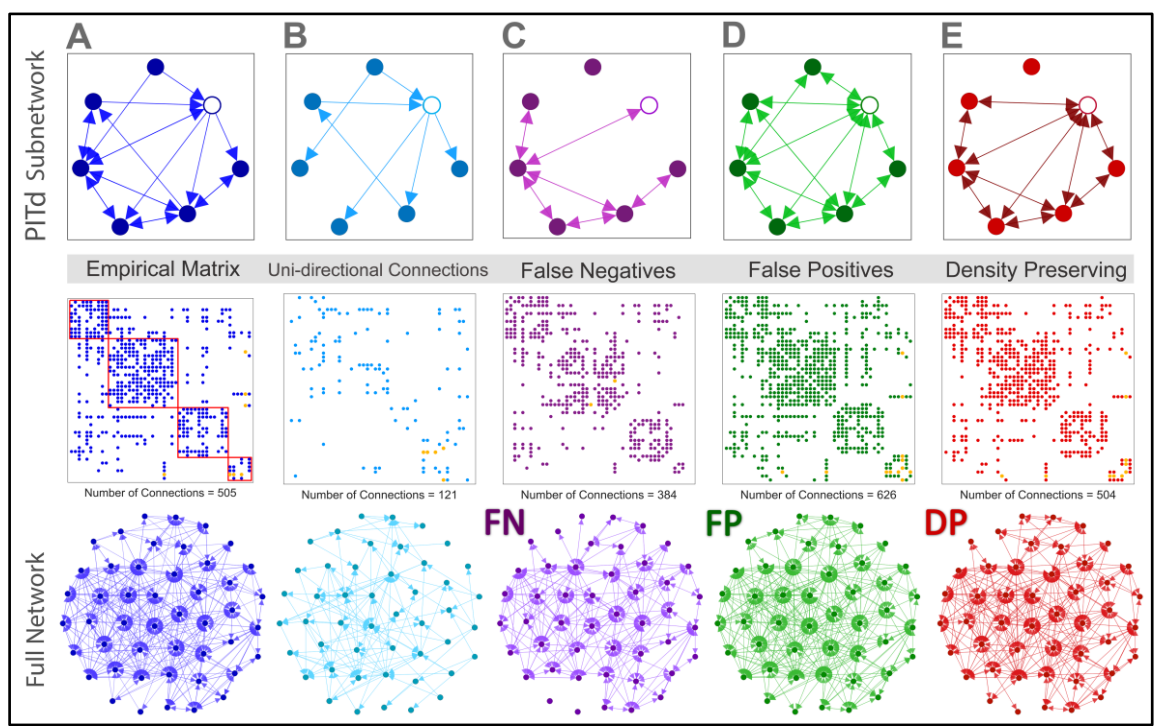

Figure 2: **Structural connectome for the macaque N=47 cortex and perturbed undirected variants, with an exemplar sub-network.** Sub-network (top) encompassing the PITd region (white node) and neighboring nodes, the adjacency matrix (middle), and the entire network (bottom) for: (A) Macaque empirical connectome with the community modules outlined in red; (B) Uni-directional connections of the connectome; (C) Connectome with uni-directional connections removed (false negative network); (D) Connectome with reciprocal connections added to uni-directional connections (false positive network); (E) Connectome with one randomly selected reciprocal connection added to a uni-directional connection for each randomly selected uni-directional connection removed (density-preserving network). In each connectome the connections linking PITd (dorsal posterior inferotemporal) to the rest of the network are colored orange.

*Network Measures*

Connectome analyses were performed using a range of common graph-theoretic network measures (Costa et al., 2007). These measures enable the quantitative comparison of connectomes across species and neuroimaging techniques while remaining computationally inexpensive (Rubinov and Sporns, 2010). Furthermore, the graphical properties of cortical systems have previously been associated with functional connectivity and evolutionary adaptations in behavior and cognition (Bullmore and Sporns, 2012, van den Heuvel et al., 2016). For each empirical connectome and associated perturbed network, we computed several graph-theoretic measures (see Supplementary Table 3), using the *Brain Connectivity Toolbox* (Rubinov and Sporns, 2010). Graph-theoretic measures for directed networks were used in all cases where applicable.

**Measures of centrality:** The *degree* of each node was calculated as the sum of the in- and out-degree, or the sum of all directed connections connecting that node to the rest of the network (Rubinov and Sporns, 2010). Network centrality identifies nodes that act as important points of information flow between regions. We used a *betweenness centrality* measure, defined as the fraction of all the shortest paths between regions that pass through a particular node (Freeman, 1978). The *participation index* or *coefficient* describes the proportion of intra- and inter-modular connections linking each node (Guimera and Amaral, 2005a). As shown in Supplementary Table 3, we used the out-participation index with the Louvain algorithm (Blondel et al., 2008) to define network modules (Rubinov and Sporns, 2010). Further details about module delineation are provided below.

**Measures of functional segregation:** We calculated the *clustering coefficient*, a measure describing the proportion of a node's neighbors that are connected to each other (Fagiolo, 2007). In undirected networks it is calculated as the probability that two connections (linking three nodes) will be closed by a third connection to form a triangle. In directed networks however, a set of three nodes can generate up to eight different triangles. The function utilized in this study, clusteringcoef_bd, (Rubinov and Sporns, 2010), takes this into account.

**Measures of functional integration:** A path is defined as a sequence of nodes and connections that represent potential routes of information flow between two brain regions. In a directed network, connections comprising a path must be arranged such that the head of one connection always precedes the tail of the subsequent connection. The *characteristic path length* for each network was calculated as the average shortest distance between all pairs of nodes (Watts and Strogatz, 1998). We also calculated the *global efficiency* of each network as the average nodal efficiency, which is the reciprocal of the harmonic mean of the shortest path length between all pairs of nodes (Latora and Marchiori, 2001).

**Small worldness:** Lastly, we measured the small-world characteristics of each network (Watts and Strogatz, 1998). For each node and for the network (see Supplementary Table 3), the *small-world index* was classified as the clustering coefficient divided by the characteristic path length of the network, with a comparison to a directed random network, makerandCIJ_dir, (Rubinov and Sporns, 2010), unless otherwise stated (Humphries and Gurney, 2008). This index combines local and global topological properties and has been linked to network efficiency (Bassett and Bullmore, 2006).

**Community detection and modularity:** We generated consensus matrices to describe the community structure of each empirical connectome (Lancichinetti and Fortunato, 2012). Specifically, 100 runs of the Louvain modularity algorithm (Blondel et al., 2008) were performed to generate a set of modular decompositions for each empirical connectome. The different runs did not necessarily yield identical decompositions due to degeneracy of the solution space and the stochastic nature of the algorithm. A consensus modularity matrix was determined for the 100 decompositions such that each element in the consensus matrix stored the proportion of runs for which a particular pair of nodes comprised the same module. The consensus modularity matrix was then thresholded (retaining values >0.4) and 100 runs of the Louvain algorithm were performed on the thresholded consensus matrix. This process was iterated until the consensus matrix converged and did not change between successive iterations. The macaque N=47 network required a greater number of iterations before a consistent community structure could be achieved (macaque N=47: 408, macaque N=71: 2, macaque N=242: 5, cat: 4, mouse: 36, C. elegans: 2).

For the perturbed networks with all uni-directional connections altered, a single consensus matrix and consistent modularity was obtained for the FN and FP networks. For the rank correlation-coefficient analyses, the modularity for each perturbed network remained the same as that assigned to the associated empirical connectome. These perturbed networks only had a small percentage of uni-directional connections altered (5%). With these measures we intended to isolate the effect of directionality on the ranking of nodes by each graph-theoretic measure, and, therefore, used the empirical consensus modularity for the (participation index) calculations on each type of perturbed network.

For DP networks with 100% of connections altered, a consensus matrix was obtained for each trial (see Supplementary Table 2 for more details). For other perturbed networks where 5%, 10% and 20% of uni-directional connections are altered, consensus modularity matrices were obtained for each run (50 runs, see Supplementary Table 2) and for each type of network (FN, FP, and DP).

*Classification of Highly Connected Regions*

Core nodes were determined using the core-periphery algorithm, function core_periphery_dir from the Brain Connectivity Toolbox (Rubinov and Sporns, 2010), with gamma=1, which subdivides all nodes in the network into either core or periphery groups of similar size. Hubs were defined as regions with a degree at least one standard deviation above the mean (Sporns et al., 2007), and super hubs were classified as those with a degree of at least 1.5 standard deviations above the mean (see Fig. 4A for an example). Super hubs were defined to evaluate the robustness of hub nodes to the progressive removal of connection directionality. More specifically, we aimed to assess whether super hubs would be demoted to hubs or non-hub nodes as directionality information was lost.

We tested the resilience of the classification of nodes belonging to the core of the network, or the set of hubs and super hubs. For each perturbed network, the accuracy of the classification of nodes into each of these three groups (core, hubs, and super hubs) was compared with the empirical connectomes. For each group, the accuracy, or matching index, $A$ was computed taking into account the number of nodes with common classification and the number of mismatched nodes that had a different classification between the empirical and the perturbed networks. More precisely, $A$ was given by the simple matching index:

$$A = \frac{C}{C+(N_e - C) + (N_b - C)}, \quad (1)$$

Where $C$ was the number of overlapping nodes within the same group between the empirical and perturbed networks; $N_e$ was the number of nodes within this group for the empirical connectome; and $N_b$ was the number of nodes within this group for the perturbed network. This measure of accuracy attained a minimum of 0 when there was no overlap between the connectomes and a maximum of 1 for a perfect overlap.

The participation index can be used to classify nodes, and has been applied to hubs (Guimera and Amaral, 2005b). Hubs with large participation index connect areas from different modules. Supplementary Table 4 lists the regions classified as hubs for each empirical network, as either connector (with a participation index Y > 0.35) or provincial (Y ≤ 0.35) hubs. Consistent with other studies (Sporns et al., 2007), node degree (as the sum of the in- and out- degree) was used to define the set of hubs based on their topological role within the network.

*Quantifying Changes in Network Measures*

To investigate changes in node-specific features between the empirical connectomes and corresponding perturbed networks, we developed a measure to quantify the change in the ranking of nodes. Nodes can be ranked with any of a number of graph-theoretic measures. The *Rank Shift Index* (*RSI*) represents the sum of the absolute value of the difference between the ranking of the empirical ($E$) and perturbed ($B$) matrices for each node, divided by the maximum possible difference ($D$) in which the ranks of the network are reversed:

$$\text{RSI} = \sum_{i=1}^{N} \frac{|E_i - B_i|}{D}, \quad (2)$$

A RSI of zero indicates no change, and an index of one indicates a complete inversion in the rank order (see Fig. 5). Node-level changes were also measured by the Spearman rank correlation (Spearman, 1904) and Kendall coefficient (Kendall, 1938).

**Results**

To understand the effects of neglecting connection directionality on the structural properties of connectomes, we compared several directed brain networks across multiple species, including three macaque connectomes (with different parcellation schemes), a cat, a mouse, and a C. elegans connectome. The characteristics of each of these networks were analyzed using a range of network measures: degree, betweenness centrality, clustering coefficient, characteristic path length, global efficiency, participation index, and small world index.

We altered uni-directional connections according to one of three schemes (see Methods) to progressively eliminate information about connection directionality. We then quantified the inaccuracies in graph-theoretic measures admitted through this loss of directionality information. We begin with the degree-preserving (DP) scheme and consider the extreme case in which all uni-directional connections are eliminated, resulting in an undirected network. In particular, we compare the network characteristics of selected regions of interest (ROIs) across the empirical connectomes and single-trial DP counterparts (Fig. 3). These ROIs (shown as the red matrix entries in Fig. 3A) occupy peripheral locations in the network topology, have low degree, and the sub-network of the local neighborhood surrounding each ROI can be clearly represented (Fig. 3B). From the empirical to the DP sub-networks, uni-directional connections are eliminated and made bidirectional, resulting in changes to graph-theoretic measures characterizing these regions. Figure 3C illustrates the relative graph-theoretic metrics at these exemplar regions for the empirical and DP sub-networks. Although the mean degree of the DP network is preserved, at the node level, the degree may increase or decrease depending on whether the uni-directional connections surrounding the node of interest received more false positive or false negative alterations. Likewise, clustering and small worldness also exhibit trial-dependent changes based on how the neighbors of these exemplar regions and the whole network topology are affected.

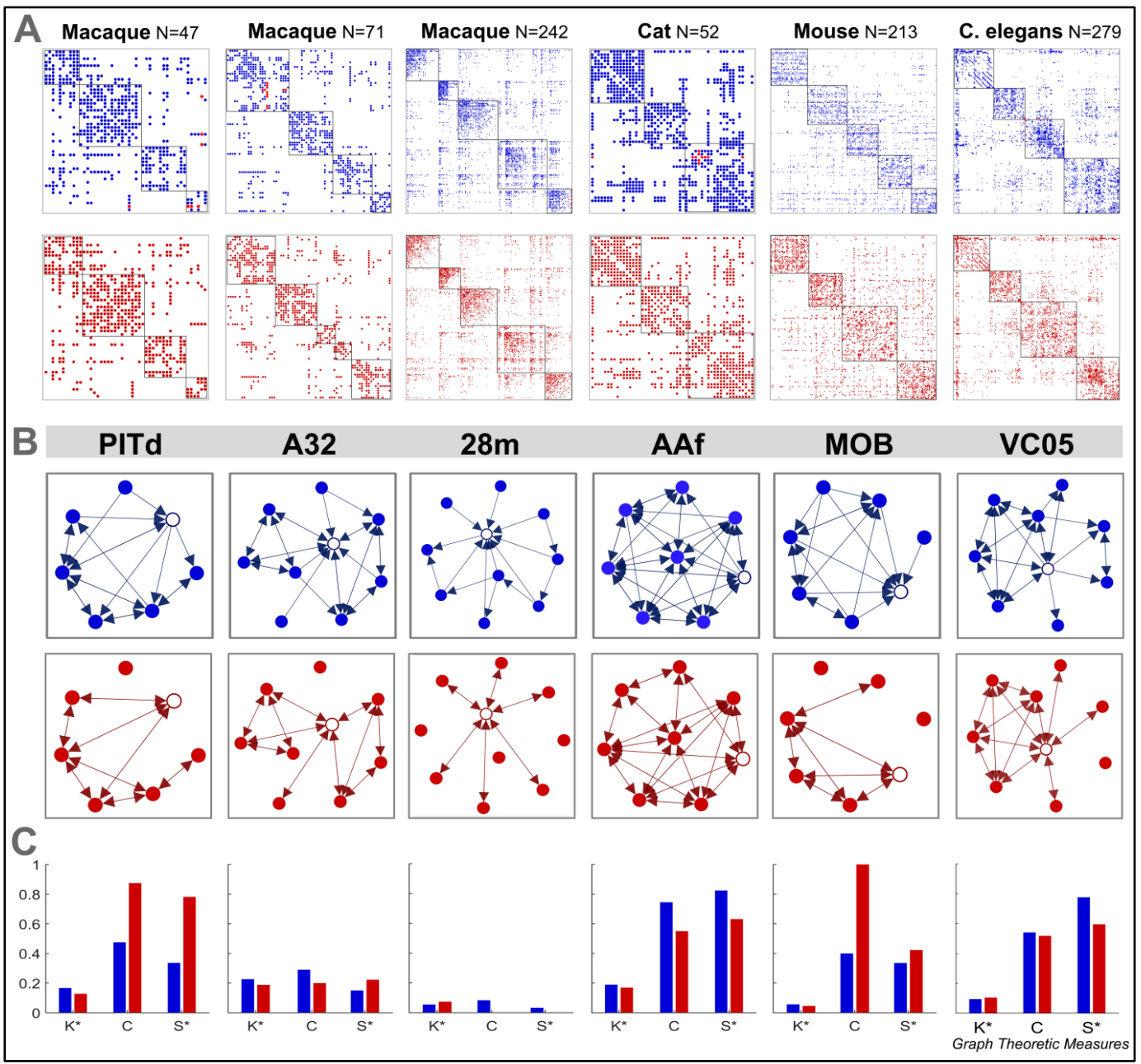

Figure 3: **Graph-theoretic measures for a specific region of interest from each empirical and density-preserving connectome.** (A) Empirical (blue) and density-preserving (red, an illustrative single trial with 100% of uni-directional connections altered) connectomes. Nodal regions are arranged into modular communities and the connections connecting the region of interest to the rest of the network in the empirical connectome are colored red. (B) Labels for each region of interest (top), and sub-networks of the local neighborhood around each region of interest (white node). (C) Graph theoretic measures at the selected brain region for the empirical and density-preserving networks. Graph-theoretic measures are as follows: K=Degree, C=Clustering coefficient and S=Small-world index ($S_i^{\rightarrow}$). *Normalized by the maximum value of that measure across all nodes in their respective network. PITd: dorsal posterior inferotemporal, A32: anterior cingulate area 32, 28m: medial entorhinal cortex, AAF: anterior auditory field, MOB: main olfactory bulb, VC05: ventral cord neuron 5.

### *Highly Connected Regions*

Connectivity across brain regions and connections is heterogeneously distributed. Hub nodes are identified as the most connected neural regions, and have enhanced importance in information integration for cognitive functions (van den Heuvel and Sporns, 2013). Hub nodes can be further classified based on their participation index as either provincial or connector hubs, depending on their level of intra- vs. inter-module connectivity (Guimera and Amaral, 2005b, Sporns et al., 2007). Provincial hubs, with a high intra-module degree and low participation index, are thought to facilitate modular segregation. Conversely, connector hubs, with a higher participation index, are thought to assist with intermodular integration (Rubinov and Sporns, 2010). When hub regions are more densely connected among themselves than to other nodes they form a 'rich club', consisting of a central but costly backbone of pathways that serves an important role in global brain communication (Colizza et al., 2006, van den Heuvel et al., 2012, Aerts et al., 2016). Hence, alterations to directionality at hub nodes influence the network activity observed in functional connectivity. But how is the identification and characteristics of these highly significant hub regions affected when directionality is modified?

Inaccuracies may be introduced to node-specific graph-theoretic measures as connection directionality information is lost. By comparing the empirical connectomes to corresponding perturbed networks with all uni-directional connections eliminated according to the DP scheme, we see that peripheral, core and hub nodes are all impacted (Fig. 4). Even the degree, a fundamental network characteristic, is affected in these perturbed networks, as shown in Fig. 4A for each cortical area in the macaque N=47 connectome. In particular, the degree of some hub and super-hub nodes falls below the threshold used for their classification in the empirical connectome. This implies that hub nodes identified based on degree can be inaccurate when directionality within the network is neglected or unknown. To further investigate this, we redefined core, hub, and super-hub nodes for each perturbed network, and calculated their accuracy according to the empirical connectome. Fig. 4B shows the percentage of nodes that retain the same classification for core, hub and super-hub nodes across all perturbed networks. We find that the estimation of core nodes from the perturbed networks were the most accurate compared to the empirical connectomes (mean=86.7%). However, the estimation of hubs and super hubs is less precise (mean=79% and 68.2% respectively). The accuracy of nodes belonging to core, hub, and super-hub was tested with paired sample t-tests and found to be significantly different. Core (including results from all connectomes and each type of perturbed network) vs. hubs $P=0.0027$, core vs. super hubs $P=0.00001$, and hubs vs. super hubs $P=0.003$. In Supplementary Fig. 2 these results are shown for each type of perturbed network and connectome separately.

A recent study in the mouse brain (Sethi et al., 2017) showed a strong correlation between the in-degree characteristics of a brain region and its resting state functional-MRI dynamics. We therefore sought to investigate in- and out-degree separately. Supplementary Fig. 3A and B display the in- and out-degree of all cortical regions in the macaque N=47 empirical connectome and perturbed networks. In this case, the delineation of hubs and super-hub nodes depends on the directed degree, and therefore a different set are identified in Figures 2A and B. However, due to the methodology for generating the perturbed networks, the resulting in- and out-degree of each node becomes equal. This is because (when 100% of uni-directional connections are altered) the only remaining connections in each case (FN, FP or DP) are represented as bidirectional and therefore, each region has the same number of incoming connections as it has outgoing connections. Previous studies in the cat connectome have found that high in-degree nodes also show (on average) a high out-degree as well. In this connectome, 66% of rich-club nodes (defined by the summed degree) had a higher in-degree than out-degree (de Reus and van den Heuvel, 2013). A comparison across the connectomes analyzed in this study (Supplementary Fig. 3C) showed that four out of six sets of hub regions had a higher mean in-degree than out-degree. The mouse connectome however, was an interesting case for which all hub regions had a much larger out-degree.

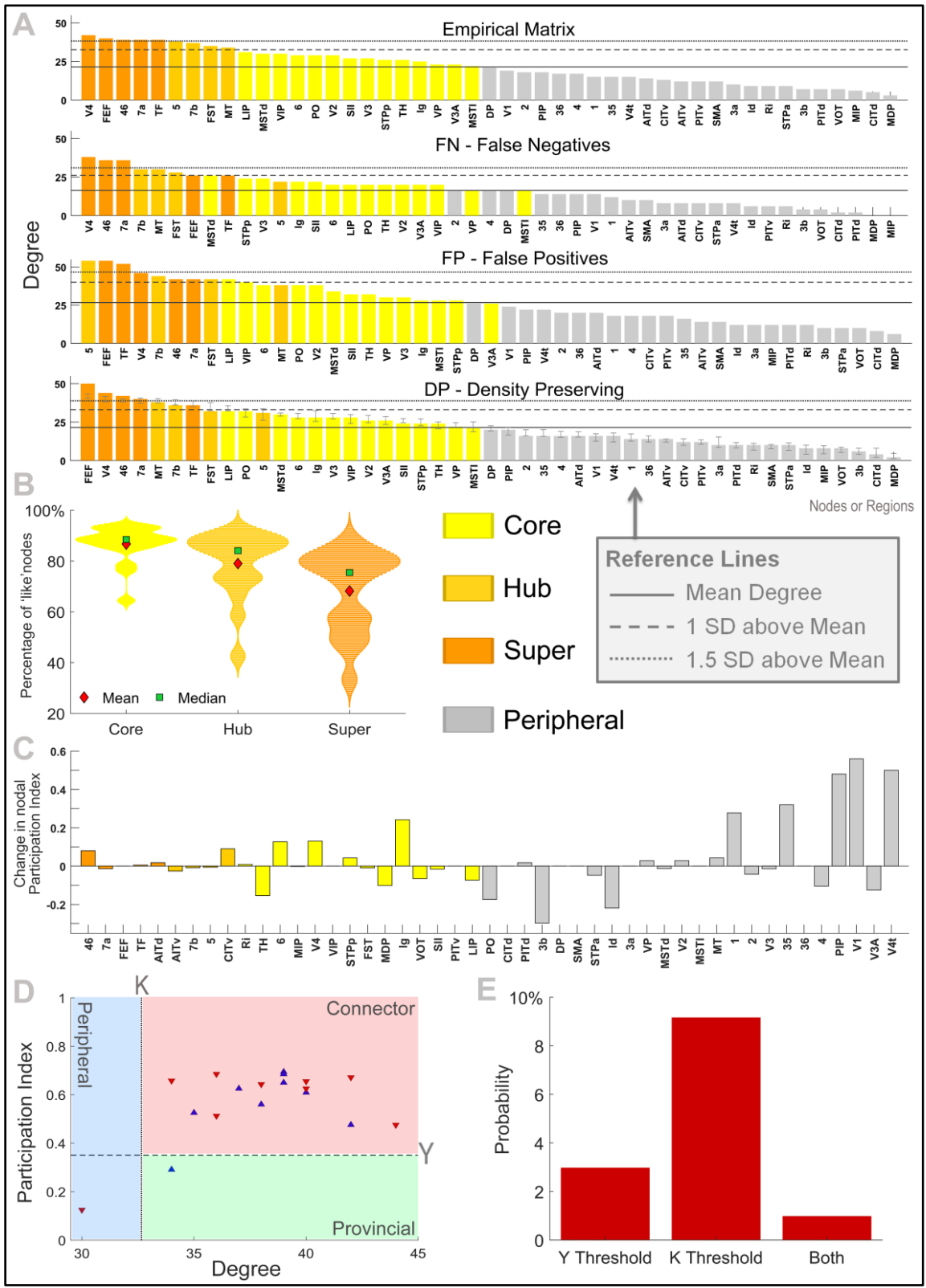

Figure 4: **Identification of hubs, changes in graph-theoretic measures at the node level, and provincial/connecter hub classification.** (A) Cortical areas of the macaque N=47 connectome sorted by degree for the empirical and each perturbed network. Hubs are defined as nodes that have a total degree (in-degree plus out-degree) one standard deviation above the mean, and super hubs are defined as nodes that have a degree 1.5 standard deviations above the mean. The density-preserving results are from an illustrative single-trial and show the standard deviation in degree for each node (over 1000 trials). (B) Percentage of core, hub, and super-hub nodes across the perturbed networks of all six connectomes that retain correct classification according to their empirical connectome (as the mean over 1000 trials). (C) Change in the participation index of each brain region from the empirical macaque N=47 connectome to an illustrative case of the density-preserving network. (D) Identification and classification of hub nodes for the empirical (blue) macaque N=47 connectome and an illustrative case of the density-preserving (red) network. The dotted line represents the hub definition based on the degree and the dashed line represents the sub-classification of hubs as either connector (Y > 0.35) or provincial (Y ≤ 0.35), based on the participation index. (E) Mean probability (across all connectomes over 1000 trials) that hub nodes will cross over either, or both of the threshold lines following density-preserving alterations in directionality, resulting in a classification that is inconsistent with the empirical connectomes. (A-E) Each perturbed network has 100% of uni-directional connections altered. Hub nodes are defined in the empirical network and remain the same in the perturbed networks.

Next, we investigate the classification of hubs based on the participation index. In comparison to peripheral regions, the participation index of hub nodes is more resilient as illustrated in Fig. 4C as the change for each region from the empirical macaque N=47 connectome to a (typical) DP example network. Because peripheral nodes have a low degree, the alterations in directionality may affect a larger proportion of these connections. Therefore, peripheral regions often show greater change in the participation index than both core and hub nodes. As illustrated in Supplementary Fig. 4, this also occurs for other graph-theoretic measures.

The relationship between participation index and degree for the set of hub nodes (defined in the empirical connectome) are displayed in Fig. 4D for the empirical macaque N=47 connectome and an illustrative DP network. Directionality alterations to the network cause changes in these measures, both of which were used to define and classify the set of hubs in the empirical connectome. As such, some of these regions in the DP network exceed the degree and participation index thresholds (degree K=1 SD above the mean and Y=0.35) resulting in misclassifications according to the empirical network. Across all connectomes, hub nodes are more likely to lose their classification based on degree, indicating that the definition of hubs based on the degree is on average 3.5 times more vulnerable to changes in directionality in comparison to the misclassification of hubs based on the participation index (Fig 4E and Supplementary Fig. 5). Supplementary Fig. 6 displays the number of core, hub, and super hubs across the connectomes (A: mean, B: individually), as defined in the empirical and each perturbed network.

*Quantifying the Errors in Node Rank when Directionality is Lost*

All the results presented thus far have pertained to perturbed networks in which all uni-directional connections are altered, yielding perturbed networks that are effectively undirected. Next, we investigate the impact of losing only a small proportion of connection directionality information. To this end, we generate perturbed networks in which the proportion of uni-directional connections altered is 5%. Changes in node-specific network measures were quantified using the rank-shift index (RSI, see Methods). This measure calculates the change in the ranking of nodes by a specific graph-theoretic measure from the empirical to the perturbed networks (see Fig. 5A). We first focus on the set of hub nodes for each connectome, finding that differences in the RSI can be seen across perturbed networks and graph-theoretic measures (Fig. 5B, super-hub results were similar). Figure 5C directly compares the effects of the FN and FP connections (perturbations) on the graph-theoretic measures, first across all nodes in the network, and then for the set of hub nodes. It can be seen that the FP connections consistently have a greater effect on the betweenness centrality and participation index, whereas the clustering coefficient and small worldness are more affected by the FN connections. For hub nodes, the RSI shows that the degree is also more affected by FP connections.

The RSI calculation is similar to the Spearman rank correlation coefficient (Spearman, 1904) and Kendall rank coefficient (Kendall, 1938) at the network level. Supplementary Fig. 7 pertains to analyses repeated with these similar, yet alternative measures and should be compared with Figs. 5B and C. Regardless of the measure used, the overall trends in the data between Fig. 5 B-C and Supplementary Fig. 7 are consistent.

Directly comparing each of the methods for altering directionality (Fig. 5D), we find that the DP networks showed the greatest RSI across almost all measures. Across connectomes the summed RSI for all graph-theoretic measures were quite similar (Fig. 5E). In particular, the mouse connectome, which has the largest proportion of uni-directional connections (see Fig. 1 and Supplementary Table 1), showed larger differences for the same percentage of altered connections.

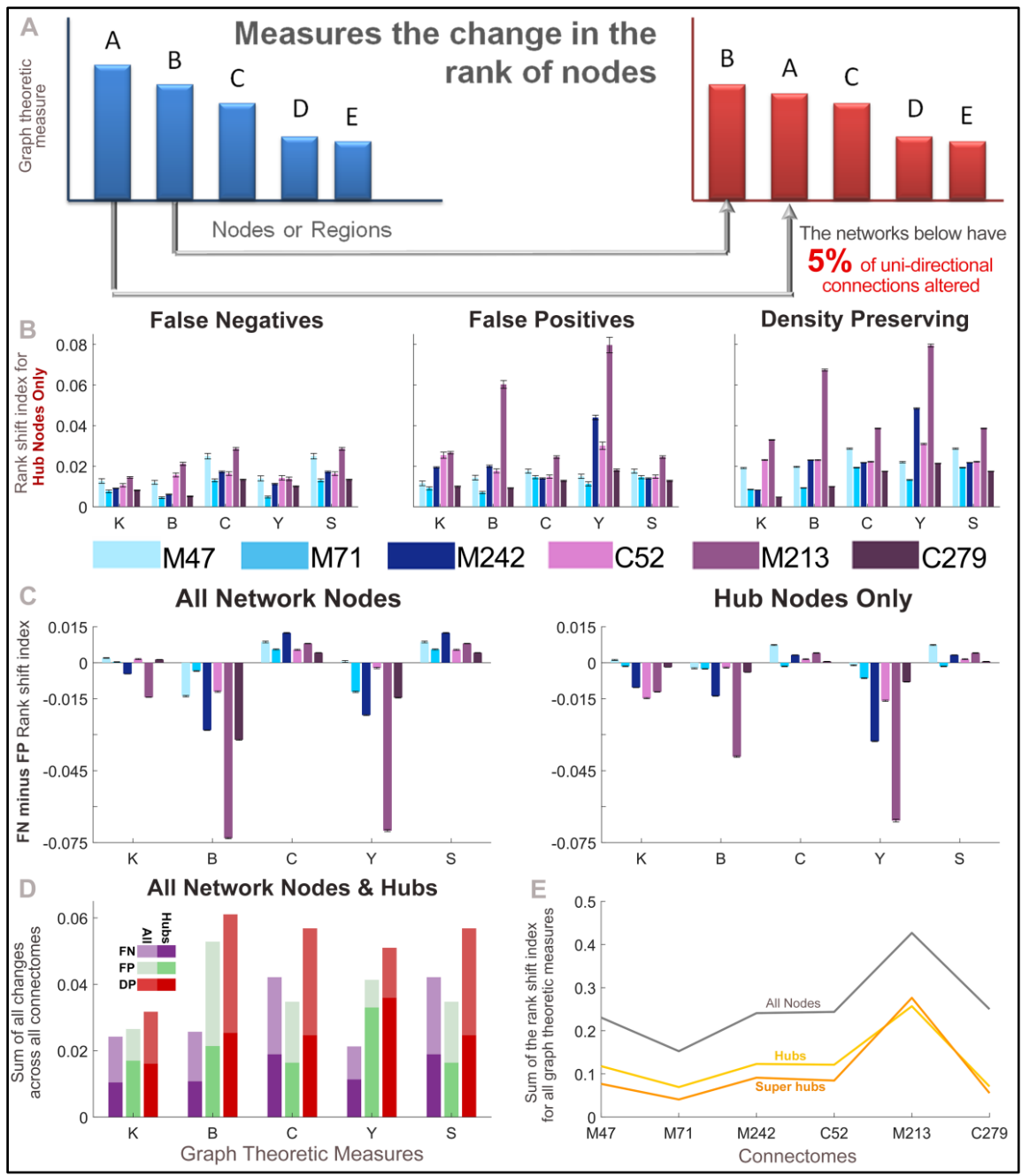

Figure 5: **Nodal changes measured by the rank-shift index.** (A) The rank-shift index quantifies the change in the rank of nodes from the empirical connectome to the perturbed network when they are ordered by a particular graph-theoretic measure. More specifically, it calculates the sum of the difference between graph-theoretic values for each node in the empirical and perturbed matrices, divided by the maximum potential difference that could exist between these two networks (where a value of 0 indicates no change, and a value of 1 indicates the maximum change). See methods for further explanation. (B) Rank-shift index of hub nodes across all perturbed networks, for each graph-theoretic measure. (C) Difference in the rank-shift index between the false-negative and false-positive networks for all nodes (left), and hub nodes (right). A positive value indicates that the false negative connections cause greater changes in the ranking of nodes, whereas a negative value indicates the same for false positive connections. (D) Rank-shift index for each graph-theoretic measure summed across all connectomes. (E) Rank-shift index values summed across all graph-theoretic measures for each density-preserving connectome. (B-E) Results correspond to the mean over 50 trials for which 5% of randomly selected uni-directional connections are modified in each perturbed network (error bars show the standard error of the mean). Graph-theoretic measures are as follows: K=Degree, B=Betweenness centrality, C=Clustering coefficient, Y=Participation index and S=Small-world index ($S_i^{\rightarrow}$). M47: the macaque connectome with 47 nodes, M71: macaque N=71, M242: macaque N=242, C52: cat, M213: mouse, C279: C. elegans.

## *Quantifying the Importance of Directed Connections in the Whole Network*

We next considered the mean changes in graph-theoretic measures in the whole network caused by the loss of directionality. We focus our analysis on perturbed networks with alterations to a small percentage of the uni-directional connections (5%, see Fig. 6). In the initial two perturbed connectomes, false negative and false positive alterations have opposite effects on network measures (Fig. 6A). The changes in betweenness (B), characteristic path length (L) and global efficiency (G) are directly dependent on the degree (K), as these connections facilitate a shorter route between nodes. The effects pertaining to clustering (C), participation index (Y), and small-world index (S) are more complex because they depend on whether the changes increase or decrease the inter-neighbor or the inter-modular connectivity. Aside from the mean degree (which is preserved in the DP networks), the effects on graph-theoretic measures were mostly similar across the FP and DP perturbed networks. To better understand the role of uni-directional connections, we next compare how false positive and false negative modifications affect the mean graph-theoretic measures of networks (Fig 6B). When it is not possible to distinguish the directionality of the connections, is it better to assume that they are bidirectional or to disregard uni-directional connections?

In the case where a subset of connections is altered, for most graph-theoretic measures the false positive uni-directional connections were more detrimental. It can be seen in Supplementary Fig. 8 that this trend remains robust as the proportion of uni-directional connections is increased (to 10% and 20%). However, the error present in each graph-theoretic measure is predictably increased. With the exception of the small worldness and degree, the FP perturbed networks consistently show the greatest changes in the mean graph-theoretic measures (Fig. 6C and Supplementary Fig. 8C, F). The participation index is the only measure directly affected by the modularity of the networks.

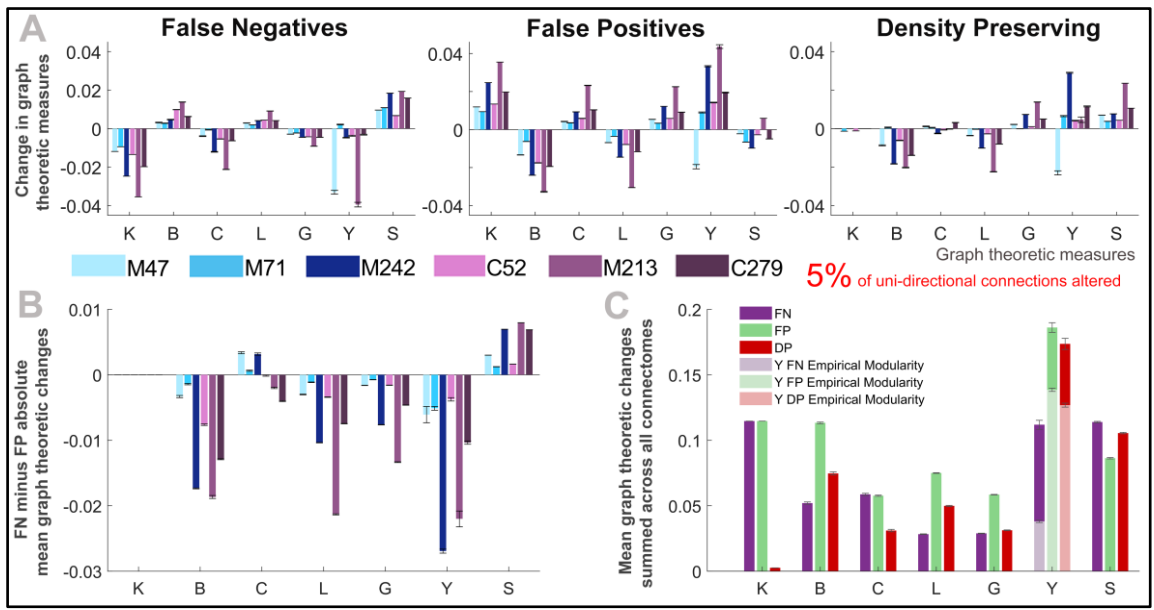

Figure 6: **Relative changes in mean graph-theoretic measures for perturbed networks.** (A) Changes in mean graph-theoretic measures across all connectomes and each type of perturbed network. (B) Difference between the changes in mean graph-theoretic measures for the false negative and false positive networks. (C) Mean changes in graph-theoretic measures for each of the perturbed networks, summed across all connectomes. Two separate modularity inputs are used the participation index calculations for the perturbed networks: the consensus modularity of the empirical networks (light colors) and the new modularity assignments for each generated perturbed network (dark colors). (A-C) All results correspond to perturbed networks with 5% of randomly selected uni-directional connections modified. The results represent the mean of these networks over 50 trials, and describe the change in the mean graph-theoretic measure (from the empirical to perturbed network) normalized by the mean of the empirical network (error bars show the standard error of the mean). Graph-theoretic measures are as follows: K=Degree, B=Betweenness centrality, C=Clustering coefficient, L=Characteristic path length, G=Global efficiency, Y=Participation index and S=Small-world index ($S^{\rightarrow}$, changes in this measure are presented as the mean over 1000 trials). M47: the macaque connectome with 47 nodes, M71: macaque N=71, M242: macaque N=242, C52: cat, M213: mouse, C279: C. elegans.

The changes in mean graph-theoretic measures are emphasized across connectomes in Supplementary Fig. 9. In the FN and FP networks, the changes for each graph-theoretic measure depend on the degree and proportion of uni-directional connections. Once again, the degree is correlated with the global efficiency and inversely correlated with the characteristic path length and betweenness. Moreover, the clustering coefficient is also correlated with the changes in degree but this is caused by the elimination of triangles from false negatives and addition of triangles from false positives.

**Discussion**

Over ten years ago, Sporns et al. (2005) proposed an influential coordinated research strategy to map the human connectome, which motivated and guided many researchers. A lot of progress has been made towards this goal with the development of diffusion-weighted imaging and tractography methods, enabling the reconstruction of several descriptions of the human connectome (Assaf and Basser, 2005, Goulas et al., 2014). However, much more research is needed to achieve an accurate, reliable and standardized representation of connectivity in the human brain. It must also be acknowledged that the methods of collation and reconstruction for these large datasets, including diffusion imaging and tract tracing, can give rise to errors and inconsistencies in the data, as discussed elsewhere (Calabrese et al., 2015, Donahue et al., 2016, Gămănuț et al., 2017). Beyond this, several parcellation schemes have been proposed for the human connectome (Cloutman and Ralph, 2012, de Reus and van den Heuvel, 2013, Honnorat et al., 2015, Glasser et al., 2016), which can each have different effects on the characterization of the network (Zalesky et al., 2010). Furthermore, the inability to resolve connection directionality noninvasively, which was originally classified as a crucial task (Sporns et al., 2005), has remained surprisingly overlooked. Without improvements in neuroimaging techniques, directionality can only be indirectly estimated for the human connectome, for example, investigating effective connectivity (Stephan et al., 2009, Friston, 2011). With current macro-scale connectome mapping techniques, connection directionality cannot be explicitly resolved.

Here, we quantified the impact of disregarding directionality in connectome analysis. Specifically, we estimated the inaccuracies in brain networks quantified by graph-theoretic measures following modifications to the uni-directional connections in connectomes of different species and parcellations.

Our analyses indicate that several network measures are susceptible to error when directionality is lost. Graph-theoretic measures are affected at both the individual-node, and network level, as is the definition of hubs. Across all networks analyzed those with a larger proportion of uni-directional connections were more extensively affected by the loss of connection directionality. This proportion is closely related to the parcellation, as finer parcellations tend to have a larger proportion of uni-directional connections. We have also compared three different schemes to generate undirected networks, which showed that the addition of reciprocal connections to a subset of existing connections (false positives) is more detrimental to graph-theoretic measures than the removal of uni-directional connections (false negatives).

*Error in the Classification of Hub Nodes*

Heterogeneity in cortical regions plays an important role in structural brain networks: Highly connected hub regions support integration of functionally and structurally segregated brain regions (van den Heuvel et al., 2012, Mišić et al., 2015, van den Heuvel et al., 2016). At these regions, neuronal dendrites have larger spine density (van den Heuvel and Sporns, 2013, Scholtens et al., 2014) and increased transcription of metabolic genes (Fulcher and Fornito, 2016). Moreover, hub nodes have high wiring cost and demand for metabolic resources, meaning their connections are more likely to become structurally damaged and symptomatic in a wide range of neuropsychiatric disorders (Crossley et al., 2014, Fornito et al., 2015, Fulcher and Fornito, 2016). For example, the increased vulnerability of hubs in Alzheimer's disease could be explained by excessive neuronal activity at these regions (Kitsak et al., 2010, de Haan et al., 2012, Raj et al., 2012). Hence, the correct identification and classification of hub regions is crucial to understanding the effects of their normal functioning (van den Heuvel and Sporns, 2013) and dysfunction (Fornito et al., 2015) within the brain network.

Our results indicate that a proportion of hubs and super-hub nodes of the human connectome are vulnerable to misclassification because the directionality of connections is not available. In particular, the classification of super-hub nodes was found to have a significant lower accuracy than hub nodes. As a caveat, we need to be aware that this measure is sensitive to noise because the number of super-hub nodes in some of the connectomes is limited.

Hubs were also classified as either connector or provincial based on their level of intra-module vs. inter-module connectivity (Guimera and Amaral, 2005b, van den Heuvel and Sporns, 2011). Previous studies have found that targeted attacks on connector hubs have a widespread effect on network dynamics due to their role in functional integration, whereas attacks on provincial hubs produce a more localized effect within communities (Honey and Sporns, 2008). It has been hypothesized that such localized damage would cause specific clinical deficits, whereas damage to connector hubs would cause complex, distributed dysfunction throughout the network (Fornito et al., 2015). We found that alterations to uni-directional connections lead to multiple errors in the classification of hub regions. Hubs were more likely to be defined incorrectly based on degree (losing their classification) rather than the participation index (changing classification between connector and provincial).

*Effect of False Positive and False Negative Connections*

Diffusion weighted and diffusion tensor imaging allow detailed reconstructions of the structural human brain network (Iturria-Medina et al., 2008, Van Essen et al., 2013). Depending on the data and specific tractography algorithms used, crossing fiber geometries can give rise to two types of errors during network reconstruction: absent connections (*false negatives)* and spurious connections (*false positives)* (Dauguet et al., 2007, Jbabdi and Johansen-Berg, 2011). These errors cannot be completely eliminated from the reconstructed network; however, when there are multiple subjects, a group threshold can be used to minimize these errors and achieve a balance between the exclusion of false positives and false negatives (de Reus and van den Heuvel, 2013, Roberts et al., 2017).

In a recent study, these two types of errors were investigated in undirected connectomes, where false negative connections were generated by pruning existing connections and false positive connections were generated by connecting pairs of unconnected nodes (Zalesky et al., 2016). False positive connections were at least twice as detrimental as false negatives to the estimation of common graph-theoretic measures: clustering coefficient, network efficiency, and modularity. This has been attributed to the modular topology of the network (Sporns and Betzel, 2016). Because nodes within the same module are likely to have a higher connection density, false negative connections were more likely to occur within modules and to be more redundant to network topology. Conversely, false positive connections were more likely to occur between modules, introducing shortcuts that have a greater impact on the graph-theoretic metrics of the network. Here we investigated the impact of perturbations to a subset of uni-directional connections, which were about half intra-modular and half inter-modular. Despite the similarity of this analysis, here we generated false negative connections by removing existing uni-directional connections and false positive connections by adding the reciprocal connections and making them bidirectional.

Our results also show that false positive connections were overall more detrimental than false negatives. This occurs for betweenness, path length, global efficiency and participation index. Notably, the small-world index and the clustering (for some connectomes) are exceptions, in which false-negative directed connections are more detrimental than false positives. For these measures, the removal of directed connections reduces the number of closed 3-node motifs in the network, which may be more detrimental. These findings suggest that graph-theoretic measures are overall more susceptible to addition of shortcuts introduced by false positive connections. A simple and immediate recommendation that follows from our results is that connectomes should be thresholded stringently to maximize specificity at the cost of sensitivity. This recommendation is very straightforward to implement and does not require the development of any new methodologies. In the mouse as well as other connectomes that have weaker uni-directional connections, a more stringent thresholding would create more false negative uni-directional connections and avoid many false positive uni-directional connections that are more detrimental for network measures. Our findings also suggest that the development of future connectome mapping methodologies should place more importance on specificity. In this way, our work can inform and guide the development of future tractography algorithms.

*Connectome Mapping and Directionality Estimation*

For the reconstruction of the macroscopic human connectome, parcellation schemes range from less than $10^2$ nodes or regions up to more than $10^5$ [see for example, Tzourio-Mazoyer et al. (2002), Salvador et al. (2005), Aleman-Gomez (2006), Hagmann et al. (2007), van den Heuvel et al. (2008), Glasser et al. (2016)]. The choice of parcellation can affect several local and global topological parameters of the network, lowering the reliability of comparisons between connectomes (Zalesky et al., 2010). The parcellation also affects the proportion of uni-directional connections, as coarser parcellations correspond to larger brain regions that are more likely to have reciprocal connections. For example, three of the connectomes can be considered coarse parcellations and have a relatively small proportion of uni-directional connections (macaque N=47, N=71 and cat connectomes). Nonetheless, even for these connectomes, the identification of hubs and their graph theoretic measures can result in inaccuracies due to loss of connection directionality.

We have used connectomes from various species and parcellations that were obtained using different techniques. These factors make it a complex task to compare and interpret some subtle features of the results across all connectomes. Nonetheless, the consistency of most results across connectomes suggests that they reflect general properties of brain networks and are largely independent from the techniques used to obtain these connectomes. Hence, they are also expected to be valid in other connectomes.

*Effect of Connectome Structure on Brain Dynamics*

Although the problem of directionality is a recurrent topic in connectomics, with few exceptions (Négyessy et al., 2008, Rosen and Louzoun, 2014), most work has focused on identifying the directionality of the interactions from the dynamics of nodes. The directionality of the interactions of nodes in motifs and networks is paramount to shaping the dynamics of systems (Bargmann and Marder, 2013). The dynamics of small circuits or network motifs can be substantially altered by subtle differences in connectivity patterns. For example, the presence of a single reciprocal connection can amplify the synchronization due to resonance (Gollo and Breakspear, 2014, Gollo et al., 2014); the presence of triangles (loops) can increase metastability (Gollo and Breakspear, 2014) or multistability (Levnajić, 2011) due to frustration. Moreover, the presence of an inhibitory feedback can cause anticipated synchronization between neurons (Matias et al., 2016) or cortical regions (Matias et al., 2014). Naturally, this susceptibility of the dynamics to structural perturbations goes beyond network motifs, affecting the dynamics of the whole network (Eguíluz et al., 2011, Hu et al., 2012, Gollo et al., 2015, Esfahani et al., 2016).

A basic and influential manner of summarizing the dynamics of brain networks corresponds to functional connectivity (Biswal et al., 1995), which captures linear correlations between pairs of regions, and are symmetric and undirected (Friston, 2011). Disambiguating the directionality of connections between pairs of cortical regions has been a priority in the field (Friston et al., 2003, Friston, 2011) as this directionality can reveal causal interaction between regions, or how they effectively interact (Friston et al., 2017). Furthermore, a number of methods have been proposed and utilized to determine the causal interactions between nodes (Friston et al., 2013), or to reconstruct the underlying network structure from the network dynamics (Stam et al., 2007, Timme, 2007, Napoletani and Sauer, 2008, Vicente et al., 2011, Friston et al., 2013, Tajima et al., 2015, Deng et al., 2016, Ching and Tam, 2017, López-Madrona et al., 2017, Wei et al., 2017). A better understanding of the relationship between directionality in network structure and dynamics may aid in determining causal interactions (Stephan et al., 2009).

At the network level, it is important to distinguish the roles of in- and out-degree in affecting brain dynamics. A recent study found strong relationships between the structural connectivity of a region and its BOLD (blood oxygen level dependent) signal dynamics (Sethi et al., 2017). Furthermore, several graph-theoretic measures showed stronger correlations to the network dynamics (resting state functional MRI) when directionality was taken into account. Brain regions receiving more input (larger in-degree) required longer integration time to process and combine all these inputs, which is consistent with the attributed function of rich-club association areas (Heeger, 2017), and also supports the notion of a hierarchy of timescales recapitulating the anatomical hierarchy of brain structure (Kiebel et al., 2008, Murray et al., 2014, Chaudhuri et al., 2015, Gollo et al., 2015, Cocchi et al., 2016, Gollo et al., 2016). Overall, these findings highlight the importance of the directionality of the structural connectivity to understand brain dynamics.

Despite intensive efforts, the structure-function relationship remains far from elucidated, and the issue of inferring directionality in undirected anatomical connectomes has yet to be addressed. Here we have focused on characterizing the effect of directionality on brain structure via graph-theoretic measures, and future work will characterize how perturbations to the directionality of connections influence network dynamics.

## Conclusions

Connectomes are inherently directed networks. The majority of non-invasive techniques for mapping connectomes are unable to resolve connection directionality, thereby yielding undirected approximations in which truly uni-directional connections are either overlooked or rendered bidirectional. We found that the inability to resolve connection directionality can introduce substantial error to the estimation of topological descriptors of brain networks, particularly with respect to the classification and identification of hubs. We analyzed the effect of progressively eliminating connection directionality information in six directed connectomes that were mapped with invasive techniques capable of resolving afferent and efferent connections (C. elegans, mouse, cat, and three macaque networks). We demonstrated that the identification of the most connected hubs is especially affected by the loss of connection directionality. We also found that the addition of reciprocal uni-directional connections (false positives) is more detrimental to the estimation of most topological measures than removal of uni-directional connections (false negatives). Our findings underscore the need for non-invasive connectome mapping techniques that can: (i) provide estimates of connection directionality; and (ii) yield relatively sparse and highly specific fiber maps that preference false negatives over false positives. Given that most topological properties have been found to be recapitulated across directed (macaque) and undirected (human) connectomes, at least qualitatively, resolving the directionality of human connectomes in the future will most likely not result in a radical re-appraisal of human brain network organization, but it will enable a more accurate characterization of the human connectome.

## Acknowledgements
We would like to sincerely thank Madeleine Flynn, QIMR Berghofer Medical Research Institute for her illustrations (Figure 1 brain/nervous system images). The authors acknowledge the support of the National Health and Medical Research Council of Australia (AZ, APP1047648; LLG, APP1110975).

## References

Achacoso T, Yamamoto W (1992) AY's Neuroanatomy of C. Elegans for Computation.

Aerts H, Fias W, Caeyenberghs K, Marinazzo D (2016) Brain networks under attack: robustness properties and the impact of lesions. Brain aww194.

Aleman-Gomez Y (2006) IBASPM: toolbox for automatic parcellation of brain structures. In: 12th Annual Meeting of the Organization for Human Brain Mapping June 11-15, 2006 Florence, Italy.

Assaf Y, Basser PJ (2005) Composite hindered and restricted model of diffusion (CHARMED) MR imaging of the human brain. Neuroimage 27:48-58.

Aydore S, Pantazis D, Leahy RM (2013) A note on the phase locking value and its properties. Neuroimage 74:231-244.

Bargmann CI, Marder E (2013) From the connectome to brain function. Nature methods 10:483-490.

Bassett DS, Bullmore E (2006) Small-world brain networks. The neuroscientist 12:512-523.

Bassett DS, Bullmore E, Verchinski BA, Mattay VS, Weinberger DR, Meyer-Lindenberg A (2008) Hierarchical organization of human cortical networks in health and schizophrenia. Journal of Neuroscience 28:9239-9248.

Betzel RF, Bassett DS (2016) Multi-scale brain networks. NeuroImage.

Bezgin G, Vakorin VA, van Opstal AJ, McIntosh AR, Bakker R (2012) Hundreds of brain maps in one atlas: registering coordinate-independent primate neuro-anatomical data to a standard brain. Neuroimage 62:67-76.

Biswal B, Zerrin Yetkin F, Haughton VM, Hyde JS (1995) Functional connectivity in the motor cortex of resting human brain using echo-planar mri. Magnetic resonance in medicine 34:537-541.

Blondel VD, Guillaume J-L, Lambiotte R, Lefebvre E (2008) Fast unfolding of communities in large networks. Journal of statistical mechanics: theory and experiment 2008:P10008.

Boussaoud D, Ungerleider LG, Desimone R (1990) Pathways for motion analysis: cortical connections of the medial superior temporal and fundus of the superior temporal visual areas in the macaque. Journal of Comparative Neurology 296:462-495.

Bullmore E, Sporns O (2012) The economy of brain network organization. Nature Reviews Neuroscience 13:336-349.

Calabrese E, Badea A, Cofer G, Qi Y, Johnson GA (2015) A diffusion MRI tractography connectome of the mouse brain and comparison with neuronal tracer data. Cerebral Cortex 25:4628-4637.

Chaudhuri R, Knoblauch K, Gariel M-A, Kennedy H, Wang X-J (2015) A large-scale circuit mechanism for hierarchical dynamical processing in the primate cortex. Neuron 88:419-431.

Ching ES, Tam H (2017) Reconstructing links in directed networks from noisy dynamics. Physical Review E 95:010301.

Cloutman LL, Ralph L (2012) Connectivity-based structural and functional parcellation of the human cortex using diffusion imaging and tractography. Frontiers in neuroanatomy 6:34.

Cocchi L, Sale MV, Gollo LL, Bell PT, Nguyen VT, Zalesky A, Breakspear M, Mattingley JB (2016) A hierarchy of timescales explains distinct effects of local inhibition of primary visual cortex and frontal eye fields. eLife 5:e15252.

Colizza V, Flammini A, Serrano MA, Vespignani A (2006) Detecting rich-club ordering in complex networks. Nature physics 2:110-115.

Costa LdF, Rodrigues FA, Travieso G, Villas Boas PR (2007) Characterization of complex networks: A survey of measurements. Advances in physics 56:167-242.

Crossley NA, Mechelli A, Scott J, Carletti F, Fox PT, McGuire P, Bullmore ET (2014) The hubs of the human connectome are generally implicated in the anatomy of brain disorders. Brain 137:2382-2395.

Dauguet J, Peled S, Berezovskii V, Delzescaux T, Warfield SK, Born R, Westin C-F (2007) Comparison of fiber tracts derived from in-vivo DTI tractography with 3D histological neural tract tracer reconstruction on a macaque brain. Neuroimage 37:530-538.

de Haan W, Mott K, van Straaten EC, Scheltens P, Stam CJ (2012) Activity dependent degeneration explains hub vulnerability in Alzheimer's disease. PLoS Comput Biol 8:e1002582.

de Reus MA, van den Heuvel MP (2013) The parcellation-based connectome: limitations and extensions. Neuroimage 80:397-404.

Deng B, Deng Y, Yu H, Guo X, Wang J (2016) Dependence of inter-neuronal effective connectivity on synchrony dynamics in neuronal network motifs. Chaos, Solitons & Fractals 82:48-59.

Donahue CJ, Sotiropoulos SN, Jbabdi S, Hernandez-Fernandez M, Behrens TE, Dyrby TB, Coalson T, Kennedy H, Knoblauch K, Van Essen DC (2016) Using diffusion tractography to predict cortical connection strength and distance: a quantitative comparison with tracers in the monkey. Journal of Neuroscience 36:6758-6770.

Durbin RM (1987) Studies on the development and organisation of the nervous system of Caenorhabditis elegans. University of Cambridge Cambridge.

Eguíluz VM, Pérez T, Borge-Holthoefer J, Arenas A (2011) Structural and functional networks in complex systems with delay. Physical Review E 83:056113.

Esfahani ZG, Gollo LL, Valizadeh A (2016) Stimulus-dependent synchronization in delayed-coupled neuronal networks. Scientific reports 6.

Fagiolo G (2007) Clustering in complex directed networks. Physical Review E 76:026107.

Felleman DJ, Van Essen DC (1991) Distributed hierarchical processing in the primate cerebral cortex. Cerebral cortex 1:1-47.

Fornito A, Zalesky A, Breakspear M (2015) The connectomics of brain disorders. Nature Reviews Neuroscience 16:159-172.

Freeman LC (1978) Centrality in social networks conceptual clarification. Social networks 1:215-239.

Friston K, Moran R, Seth AK (2013) Analysing connectivity with Granger causality and dynamic causal modelling. Current opinion in neurobiology 23:172-178.

Friston K, Preller KH, Mathys C, Cagnan H, Heinzle J, Razi A, Zeidman P (2017) Dynamic causal modelling revisited. NeuroImage.

Friston KJ (2011) Functional and effective connectivity: a review. Brain connectivity 1:13-36.

Friston KJ, Harrison L, Penny W (2003) Dynamic causal modelling. Neuroimage 19:1273-1302.

Fulcher BD, Fornito A (2016) A transcriptional signature of hub connectivity in the mouse connectome. Proceedings of the National Academy of Sciences 113:1435-1440.

Gămănuț R, Kennedy H, Toroczkai Z, Van Essen D, Knoblauch K, Burkhalter A (2017) The Mouse Cortical Interareal Network Reveals Well Defined Connectivity Profiles and an Ultra Dense Cortical Graph. bioRxiv 156976.

Garlaschelli D, Loffredo MI (2004) Patterns of link reciprocity in directed networks. Physical review letters 93:268701.

Glasser MF, Coalson TS, Robinson EC, Hacker CD, Harwell J, Yacoub E, Ugurbil K, Andersson J, Beckmann CF, Jenkinson M (2016) A multi-modal parcellation of human cerebral cortex. Nature.

Gollo LL, Breakspear M (2014) The frustrated brain: from dynamics on motifs to communities and networks. Phil Trans R Soc B 369:20130532.

Gollo LL, Mirasso C, Sporns O, Breakspear M (2014) Mechanisms of zero-lag synchronization in cortical motifs. PLoS Comput Biol 10:e1003548.

Gollo LL, Roberts JA, Cocchi L (2016) Mapping how local perturbations influence systems-level brain dynamics. arXiv preprint arXiv:160900491.

Gollo LL, Zalesky A, Hutchison RM, van den Heuvel M, Breakspear M (2015) Dwelling quietly in the rich club: brain network determinants of slow cortical fluctuations. Phil Trans R Soc B 370:20140165.

Goulas A, Bastiani M, Bezgin G, Uylings HB, Roebroeck A, Stiers P (2014) Comparative analysis of the macroscale structural connectivity in the macaque and human brain. PLoS Comput Biol 10:e1003529.

Guimera R, Amaral LAN (2005a) Cartography of complex networks: modules and universal roles. Journal of Statistical Mechanics: Theory and Experiment 2005:P02001.

Guimera R, Amaral LAN (2005b) Functional cartography of complex metabolic networks. Nature 433:895-900.

Hagmann P, Cammoun L, Gigandet X, Meuli R, Honey CJ, Wedeen VJ, Sporns O (2008) Mapping the structural core of human cerebral cortex. PLoS Biol 6:e159.

Hagmann P, Kurant M, Gigandet X, Thiran P, Wedeen VJ, Meuli R, Thiran J-P (2007) Mapping human whole-brain structural networks with diffusion MRI. PLoS one 2:e597.

Hall DH, Russell RL (1991) The posterior nervous system of the nematode Caenorhabditis elegans: serial reconstruction of identified neurons and complete pattern of synaptic interactions. Journal of Neuroscience 11:1-22.

Harriger L, van den Heuvel MP, Sporns O (2012) Rich club organization of macaque cerebral cortex and its role in network communication. PloS one 7:e46497.

Heeger DJ (2017) Theory of cortical function. Proceedings of the National Academy of Sciences 114:1773-1782.

Honey CJ, Kötter R, Breakspear M, Sporns O (2007) Network structure of cerebral cortex shapes functional connectivity on multiple time scales. Proceedings of the National Academy of Sciences 104:10240-10245.

Honey CJ, Sporns O (2008) Dynamical consequences of lesions in cortical networks. Human brain mapping 29:802-809.

Honnorat N, Eavani H, Satterthwaite TD, Gur RE, Gur RC, Davatzikos C (2015) GraSP: geodesic graph-based segmentation with shape priors for the functional parcellation of the cortex. Neuroimage 106:207-221.

Hu Y, Trousdale J, Josić K, Shea-Brown E (2012) Motif statistics and spike correlations in neuronal networks. BMC Neuroscience 13:P43.

Humphries MD, Gurney K (2008) Network 'small-world-ness': a quantitative method for determining canonical network equivalence. PloS one 3:e0002051.

Iturria-Medina Y, Sotero RC, Canales-Rodríguez EJ, Alemán-Gómez Y, Melie-García L (2008) Studying the human brain anatomical network via diffusion-weighted MRI and Graph Theory. Neuroimage 40:1064-1076.

Jbabdi S, Johansen-Berg H (2011) Tractography: where do we go from here? Brain connectivity 1:169-183.

Kandel ER, Schwartz JH, Jessell TM, Siegelbaum SA, Hudspeth A (2000) Principles of neural science: McGraw-hill New York.

Kendall MG (1938) A new measure of rank correlation. Biometrika 30:81-93.

Kiebel SJ, Daunizeau J, Friston KJ (2008) A hierarchy of time-scales and the brain. PLoS Comput Biol 4:e1000209.

Kitsak M, Gallos LK, Havlin S, Liljeros F, Muchnik L, Stanley HE, Makse HA (2010) Identification of influential spreaders in complex networks. Nature physics 6:888-893.

Kötter R (2004) Online retrieval, processing, and visualization of primate connectivity data from the CoCoMac database. Neuroinformatics 2:127-144.

Kötter R, Wanke E (2005) Mapping brains without coordinates. Philosophical Transactions of the Royal Society of London B: Biological Sciences 360:751-766.

Lancichinetti A, Fortunato S (2012) Consensus clustering in complex networks. Scientific reports 2:336.

Latora V, Marchiori M (2001) Efficient behavior of small-world networks. Physical review letters 87:198701.

Levnajić Z (2011) Emergent multistability and frustration in phase-repulsive networks of oscillators. Physical Review E 84:016231.

López-Madrona VJ, Matias FS, Pereda E, Canals S, Mirasso CR (2017) On the role of the entorhinal cortex in the effective connectivity of the hippocampal formation. Chaos: An Interdisciplinary Journal of Nonlinear Science 27:047401.

Maier-Hein KH, Neher PF, Houde J-C, Côté M-A, Garyfallidis E, Zhong J, Chamberland M, Yeh F-C, Lin Y-C, Ji Q (2017) The challenge of mapping the human connectome based on diffusion tractography. Nature Communications 8:1349.

Markov NT, Ercsey-Ravasz M, Ribeiro Gomes A, Lamy C, Magrou L, Vezoli J, Misery P, Falchier A, Quilodran R, Gariel M (2012) A weighted and directed interareal connectivity matrix for macaque cerebral cortex. Cerebral cortex 24:17-36.

Matias FS, Gollo LL, Carelli PV, Bressler SL, Copelli M, Mirasso CR (2014) Modeling positive Granger causality and negative phase lag between cortical areas. NeuroImage 99:411-418.

Matias FS, Gollo LL, Carelli PV, Mirasso CR, Copelli M (2016) Inhibitory loop robustly induces anticipated synchronization in neuronal microcircuits. Physical Review E 94:042411.

Medaglia JD, Bassett DS (2017) Network Analyses and Nervous System Disorders. arXiv preprint arXiv:170101101.

Mišić B, Betzel RF, Nematzadeh A, Goñi J, Griffa A, Hagmann P, Flammini A, Ahn Y-Y, Sporns O (2015) Cooperative and competitive spreading dynamics on the human connectome. Neuron 86:1518-1529.

Modha DS, Singh R (2010) Network architecture of the long-distance pathways in the macaque brain. Proceedings of the National Academy of Sciences 107:13485-13490.

Murray JD, Bernacchia A, Freedman DJ, Romo R, Wallis JD, Cai X, Padoa-Schioppa C, Pasternak T, Seo H, Lee D (2014) A hierarchy of intrinsic timescales across primate cortex. Nature neuroscience 17:1661-1663.

Napoletani D, Sauer TD (2008) Reconstructing the topology of sparsely connected dynamical networks. Physical Review E 77:026103.

Négyessy L, Nepusz T, Zalányi L, Bazsó F (2008) Convergence and divergence are mostly reciprocated properties of the connections in the network of cortical areas. Proceedings of the Royal Society of London B: Biological Sciences 275:2403-2410.

Newman ME (2004) Coauthorship networks and patterns of scientific collaboration. Proceedings of the national academy of sciences 101:5200-5205.

Oh SW, Harris JA, Ng L, Winslow B, Cain N, Mihalas S, Wang Q, Lau C, Kuan L, Henry AM (2014) A mesoscale connectome of the mouse brain. Nature 508:207-214.

Raj A, Kuceyeski A, Weiner M (2012) A network diffusion model of disease progression in dementia. Neuron 73:1204-1215.

Reinoso-Suarez F (1984) Connectional patterns in parietotemporooccipital association cortex of the feline cerebral cortex. Cortical integration: Basic archicortical and cortical association levels of neural integration IBRO monograph series Val 11:255-278.

Roberts JA, Perry A, Roberts G, Mitchell PB, Breakspear M (2017) Consistency-based thresholding of the human connectome. Neuroimage 145:118-129.

Rosen Y, Louzoun Y (2014) Directionality of real world networks as predicted by path length in directed and undirected graphs. Physica A: Statistical Mechanics and its Applications 401:118-129.

Rubinov M, Sporns O (2010) Complex network measures of brain connectivity: uses and interpretations. Neuroimage 52:1059-1069.

Salvador R, Suckling J, Schwarzbauer C, Bullmore E (2005) Undirected graphs of frequency-dependent functional connectivity in whole brain networks. Philosophical Transactions of the Royal Society of London B: Biological Sciences 360:937-946.

Scannell J, Burns G, Hilgetag C, O'Neil M, Young MP (1999) The connectional organization of the cortico-thalamic system of the cat. Cerebral Cortex 9:277-299.

Scannell JW, Blakemore C, Young MP (1995) Analysis of connectivity in the cat cerebral cortex. Journal of Neuroscience 15:1463-1483.

Scholtens LH, Schmidt R, de Reus MA, van den Heuvel MP (2014) Linking macroscale graph analytical organization to microscale neuroarchitectonics in the macaque connectome. Journal of Neuroscience 34:12192-12205.

Serrano MÁ, Boguñá M, Vespignani A (2009) Extracting the multiscale backbone of complex weighted networks. Proceedings of the national academy of sciences 106:6483-6488.

Sethi SS, Zerbi V, Wenderoth N, Fornito A, Fulcher BD (2017) Structural connectome topology relates to regional BOLD signal dynamics in the mouse brain. Chaos: An Interdisciplinary Journal of Nonlinear Science 27:047405.

Shih C-T, Sporns O, Yuan S-L, Su T-S, Lin Y-J, Chuang C-C, Wang T-Y, Lo C-C, Greenspan RJ, Chiang A-S (2015) Connectomics-based analysis of information flow in the Drosophila brain. Current Biology 25:1249-1258.

Spearman C (1904) The proof and measurement of association between two things. The American journal of psychology 15:72-101.

Sporns O (2011) The human connectome: a complex network. Annals of the New York Academy of Sciences 1224:109-125.

Sporns O, Betzel RF (2016) Modular brain networks. Annual review of psychology 67:613-640.

Sporns O, Honey CJ, Kötter R (2007) Identification and classification of hubs in brain networks. PloS one 2:e1049.

Sporns O, Tononi G, Kötter R (2005) The human connectome: a structural description of the human brain. PLoS Comput Biol 1:e42.

Stam CJ, Nolte G, Daffertshofer A (2007) Phase lag index: assessment of functional connectivity from multi channel EEG and MEG with diminished bias from common sources. Human brain mapping 28:1178-1193.

Stephan KE, Kamper L, Bozkurt A, Burns GA, Young MP, Kötter R (2001) Advanced database methodology for the Collation of Connectivity data on the Macaque brain (CoCoMac). Philosophical Transactions of the Royal Society of London B: Biological Sciences 356:1159-1186.

Stephan KE, Tittgemeyer M, Knösche TR, Moran RJ, Friston KJ (2009) Tractography-based priors for dynamic causal models. Neuroimage 47:1628-1638.

Stephan KE, Zilles K, Kötter R (2000) Coordinate–independent mapping of structural and functional data by objective relational transformation (ORT). Philosophical Transactions of the Royal Society of London B: Biological Sciences 355:37-54.

Tajima S, Yanagawa T, Fujii N, Toyoizumi T (2015) Untangling brain-wide dynamics in consciousness by cross-embedding. PLoS Comput Biol 11:e1004537.

Timme M (2007) Revealing network connectivity from response dynamics. Physical review letters 98:224101.

Tournier J, Calamante F, Connelly A (2012) MRtrix: diffusion tractography in crossing fiber regions. International Journal of Imaging Systems and Technology 22:53-66.

Towlson EK, Vértes PE, Ahnert SE, Schafer WR, Bullmore ET (2013) The rich club of the C. elegans neuronal connectome. Journal of Neuroscience 33:6380-6387.

Tzourio-Mazoyer N, Landeau B, Papathanassiou D, Crivello F, Etard O, Delcroix N, Mazoyer B, Joliot M (2002) Automated anatomical labeling of activations in SPM using a macroscopic anatomical parcellation of the MNI MRI single-subject brain. Neuroimage 15:273-289.

van den Heuvel MP, Bullmore ET, Sporns O (2016) Comparative connectomics. Trends in cognitive sciences 20:345-361.

van den Heuvel MP, Kahn RS, Goñi J, Sporns O (2012) High-cost, high-capacity backbone for global brain communication. Proceedings of the National Academy of Sciences 109:11372-11377.

van den Heuvel MP, Sporns O (2011) Rich-club organization of the human connectome. The Journal of neuroscience 31:15775-15786.

van den Heuvel MP, Sporns O (2013) Network hubs in the human brain. Trends in cognitive sciences 17:683-696.

van den Heuvel MP, Stam CJ, Boersma M, Pol HH (2008) Small-world and scale-free organization of voxel-based resting-state functional connectivity in the human brain. Neuroimage 43:528-539.

Van Essen DC, Smith SM, Barch DM, Behrens TE, Yacoub E, Ugurbil K, Consortium W-MH (2013) The WU-Minn human connectome project: an overview. Neuroimage 80:62-79.

Varshney LR, Chen BL, Paniagua E, Hall DH, Chklovskii DB (2011) Structural properties of the Caenorhabditis elegans neuronal network. PLoS Comput Biol 7:e1001066.

Vicente R, Wibral M, Lindner M, Pipa G (2011) Transfer entropy—a model-free measure of effective connectivity for the neurosciences. Journal of computational neuroscience 30:45-67.

Watts DJ, Strogatz SH (1998) Collective dynamics of 'small-world' networks. nature 393:440-442.

Wei Y, Liao X, Yan C, He Y, Xia M (2017) Identifying topological motif patterns of human brain functional networks. Human Brain Mapping 38:2734-2750.

White JG, Southgate E, Thomson JN, Brenner S (1976) The structure of the ventral nerve cord of Caenorhabditis elegans. Philosophical Transactions of the Royal Society of London B: Biological Sciences 275:327-348.

White JG, Southgate E, Thomson JN, Brenner S (1986) The structure of the nervous system of the nematode Caenorhabditis elegans. Philos Trans R Soc Lond B Biol Sci 314:1-340.

Yeterian EH, Pandya DN (1985) Corticothalamic connections of the posterior parietal cortex in the rhesus monkey. Journal of Comparative Neurology 237:408-426.

Young MP (1993) The organization of neural systems in the primate cerebral cortex. Proceedings of the Royal Society of London B: Biological Sciences 252:13-18.

Ypma RJ, Bullmore ET (2016) Statistical analysis of tract-tracing experiments demonstrates a dense, complex cortical network in the mouse. PLoS computational biology 12:e1005104.

Zalesky A, Fornito A, Cocchi L, Gollo LL, van den Heuvel MP, Breakspear M (2016) Connectome sensitivity or specificity: which is more important? NeuroImage.

Zalesky A, Fornito A, Harding IH, Cocchi L, Yücel M, Pantelis C, Bullmore ET (2010) Whole-brain anatomical networks: does the choice of nodes matter? Neuroimage 50:970-983.

## SUPPLEMENTARY MATERIAL

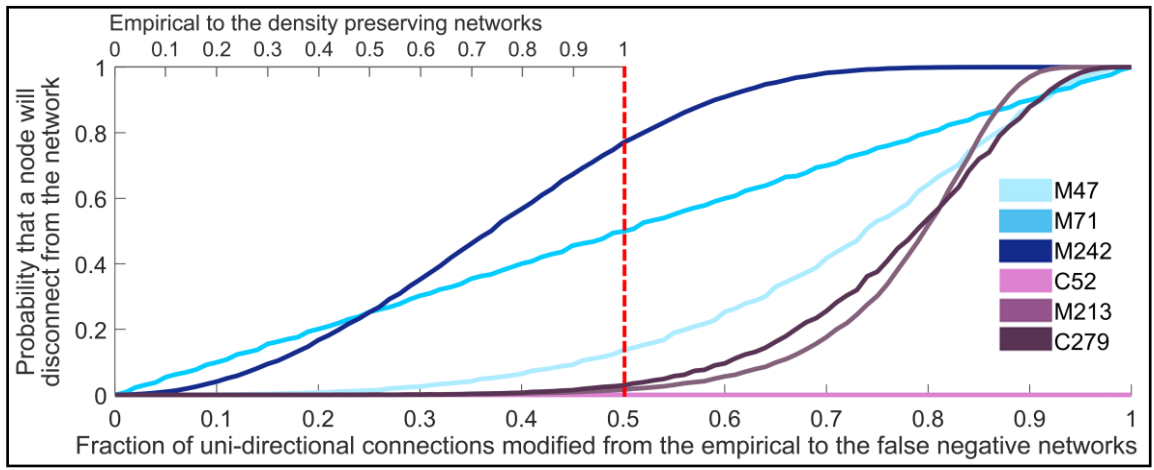

Supplementary Figure 1: **Probability that a node will disconnect from the density preserving and false negative networks.** Probability that a node will disconnect from the network as uni-directional connections are removed in the false negative perturbed networks (lower x-axis). The probability of nodal disconnection for the density preserving networks (top x-axis prior to the red dashed line) as randomly selected uni-directional connections are removed, with a reciprocal connection added to another uni-directional connection with each modification. These results describe the mean over 1000 trials. M47: the macaque connectome with 47 nodes, M71: macaque N=71, M242: macaque N=242, C52: cat, M213: mouse, C279: C. elegans.

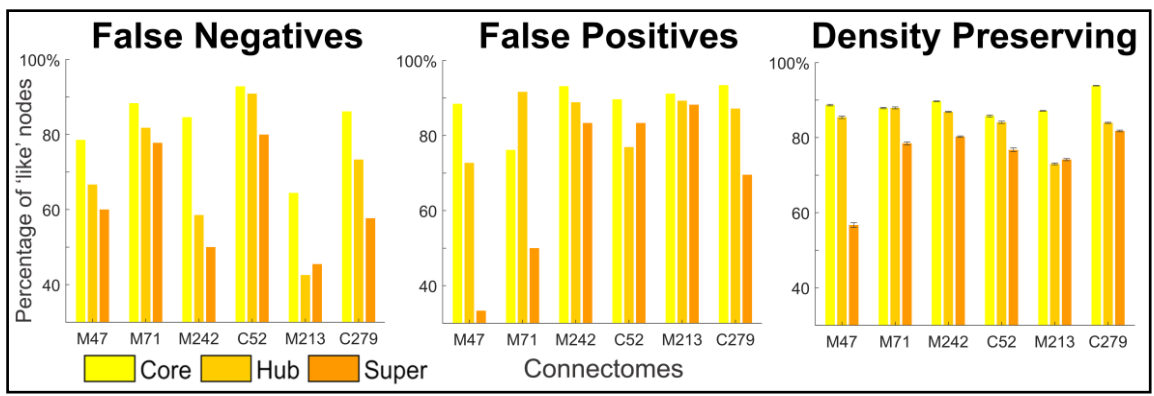

Supplementary Figure 2: **Percentage of high degree nodes classified correctly as defined by the empirical network.** These figures display the percentage of core, hub, and super-hub nodes across all connectomes that are accurate (see methods) according to those in the empirical networks when these nodes are redefined for each of the perturbed connectomes (with 100% of uni-directional connections altered; the density-preserving results show a mean over 1000 trials with the standard deviation). M47: the macaque connectome with 47 nodes, M71: macaque N=71, M242: macaque N=242, C52: cat, M213: mouse, C279: C. elegans.

Supplementary Figure 3: **Directionality effects on the in- and out- degree.** (A-B) Cortical areas of the macaque N=47 connectome sorted by in-degree (A) or out-degree (B) for the empirical and each perturbed network. Hubs are defined as nodes that have a total in-degree (A) or out-degree (B) one standard deviation above the mean, and super-hubs are defined as nodes that have an in-/out-degree 1.5 standard deviations above the mean (the density-preserving results are from an illustrative single trial). (C) The mean of the in-degree minus the out-degree for the set of hub nodes in each connectome. M47: the macaque connectome with 47 nodes, M71: macaque N=71, M242: macaque N=242, C52: cat, M213: mouse, C279: C. elegans.

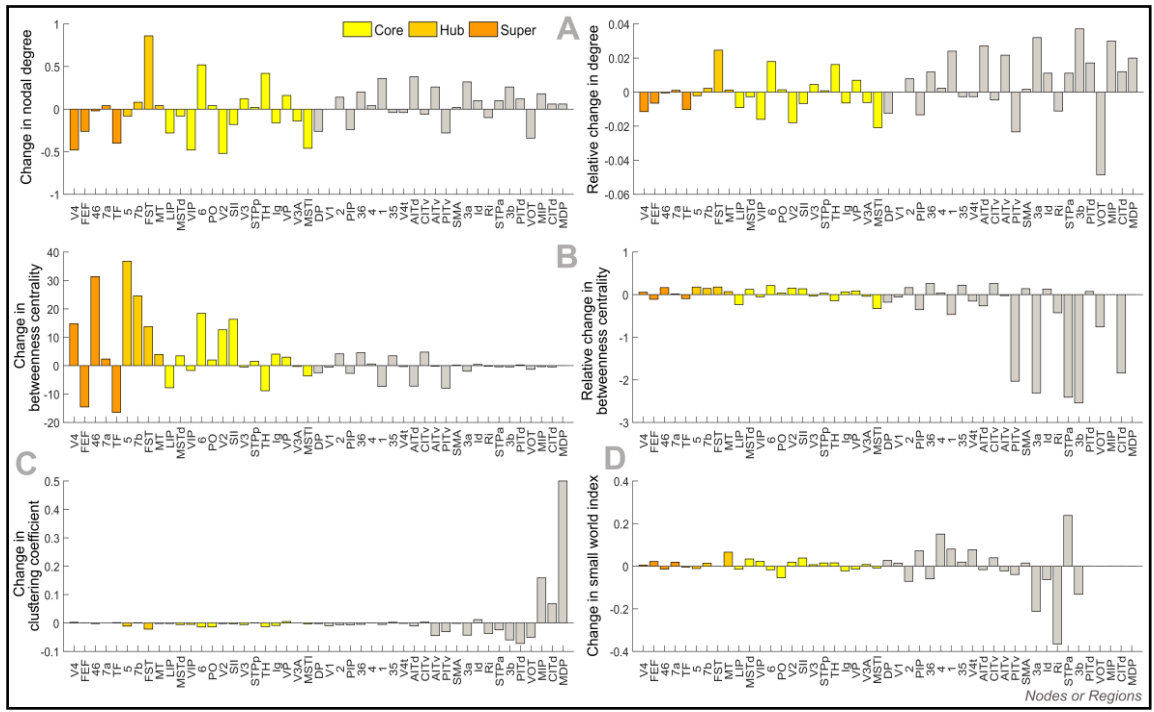

Supplementary Figure 4: **Nodal variation in graph-theoretic measures for each region in the macaque N=47 connectome from the empirical to the density preserving network.** (A) Change in the degree of each region (left) from the empirical to the density preserving network, and the relative change (right). (B) Change in the betweenness centrality of each node (left) and the relative change (right). (C) Change in the clustering coefficient of each node. (D) The change in the small-world index of each node ($S_i^{\rightarrow}$). These results were generated from the mean change across 50 trial density preserving networks, and were relative to the maximum graph-theoretic value (for each measure and region) across the macaque empirical network.

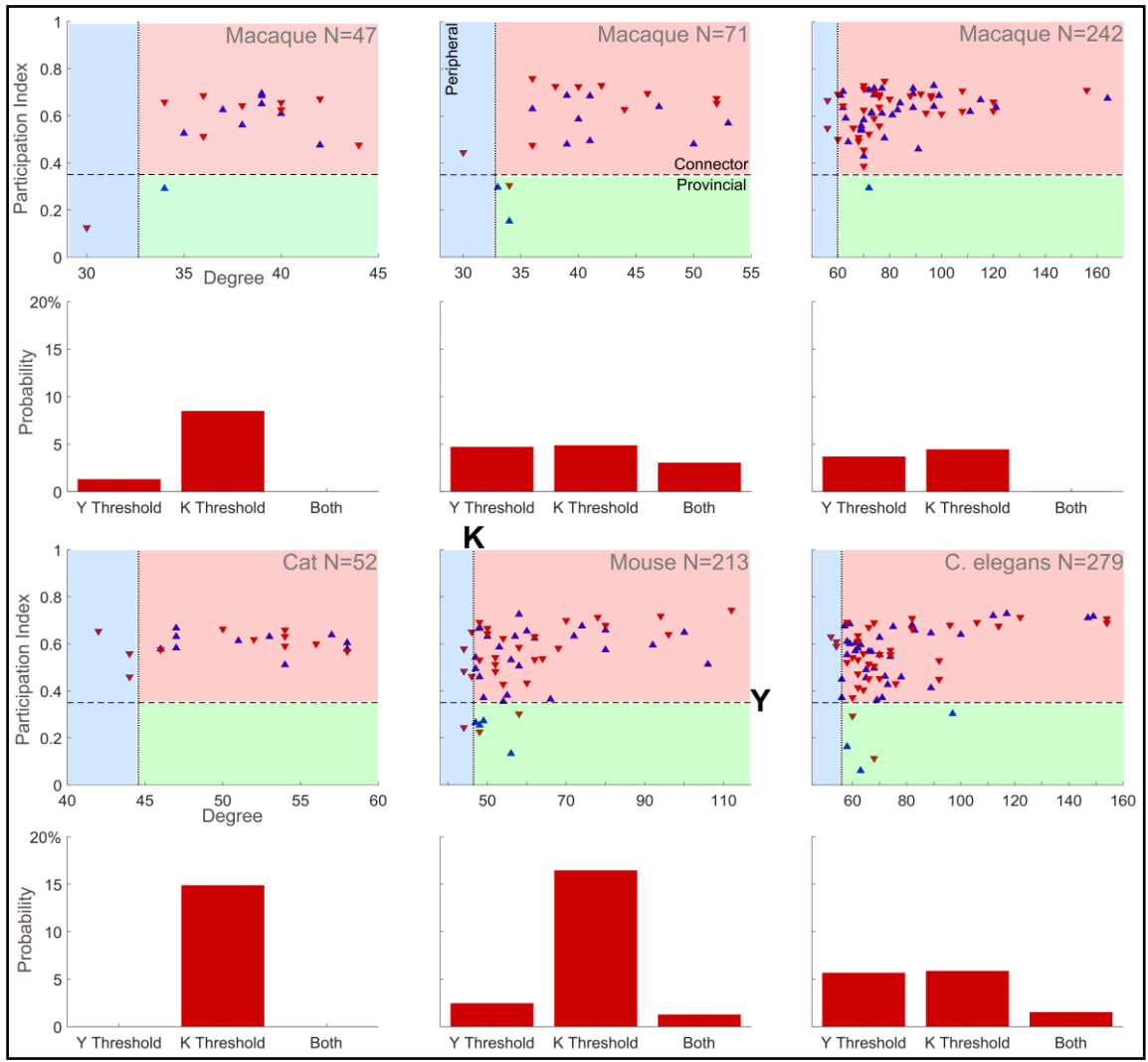

Supplementary Figure 5: **Relationship between the participation index and degree for hub nodes in each empirical and density preserving connectome.** Plots (participation vs. degree) displaying the hub nodes of the empirical (blue) network, and the same regions in the density preserving (red) networks for each connectome (taken from a single illustrative trial). The dotted line represents the hub definition based on the degree (K > one standard deviation above the mean), and the dashed line represents the further classification of hubs based on the participation index (connector hubs Y > 0.35 and provincial hubs Y ≤ 0.35). The bar figures below show the probability that a hub node will cross over either, or both of the threshold lines following density-preserving alterations in directionality (over 1000 trials), resulting in a classification that is inconsistent with the empirical connectome.

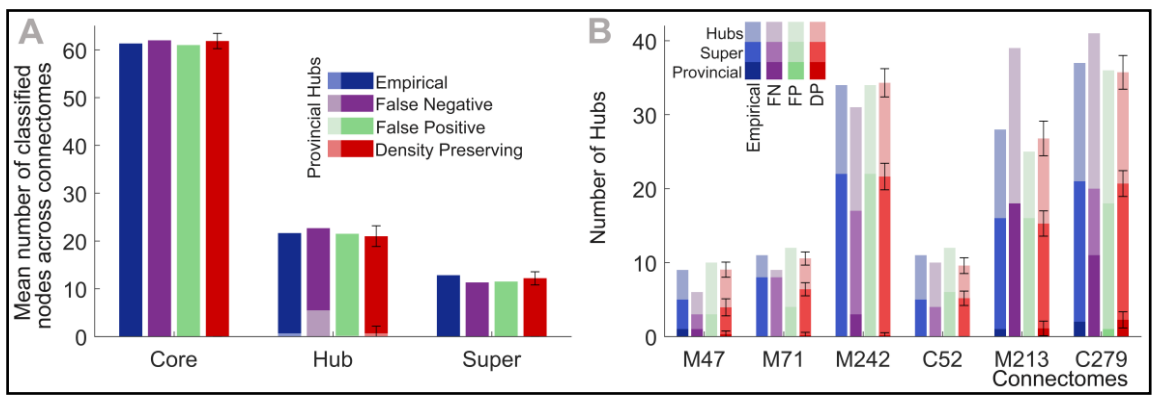

Supplementary Figure 6: **Number of core, hub and super-hub nodes identified in each empirical and perturbed network across all connectomes.** (A) Mean number of each type of highly connected region across all connectomes, for the empirical and each perturbed network (100% of uni-directional connections altered). (B) Number of each type of hub region across all connectomes, for the empirical and each perturbed network. Results for the density preserving network are the mean of 50 trials and show the standard deviation. M47: the macaque connectome with 47 nodes, M71: macaque N=71, M242: macaque N=242, C52: cat, M213: mouse, C279: C. elegans.

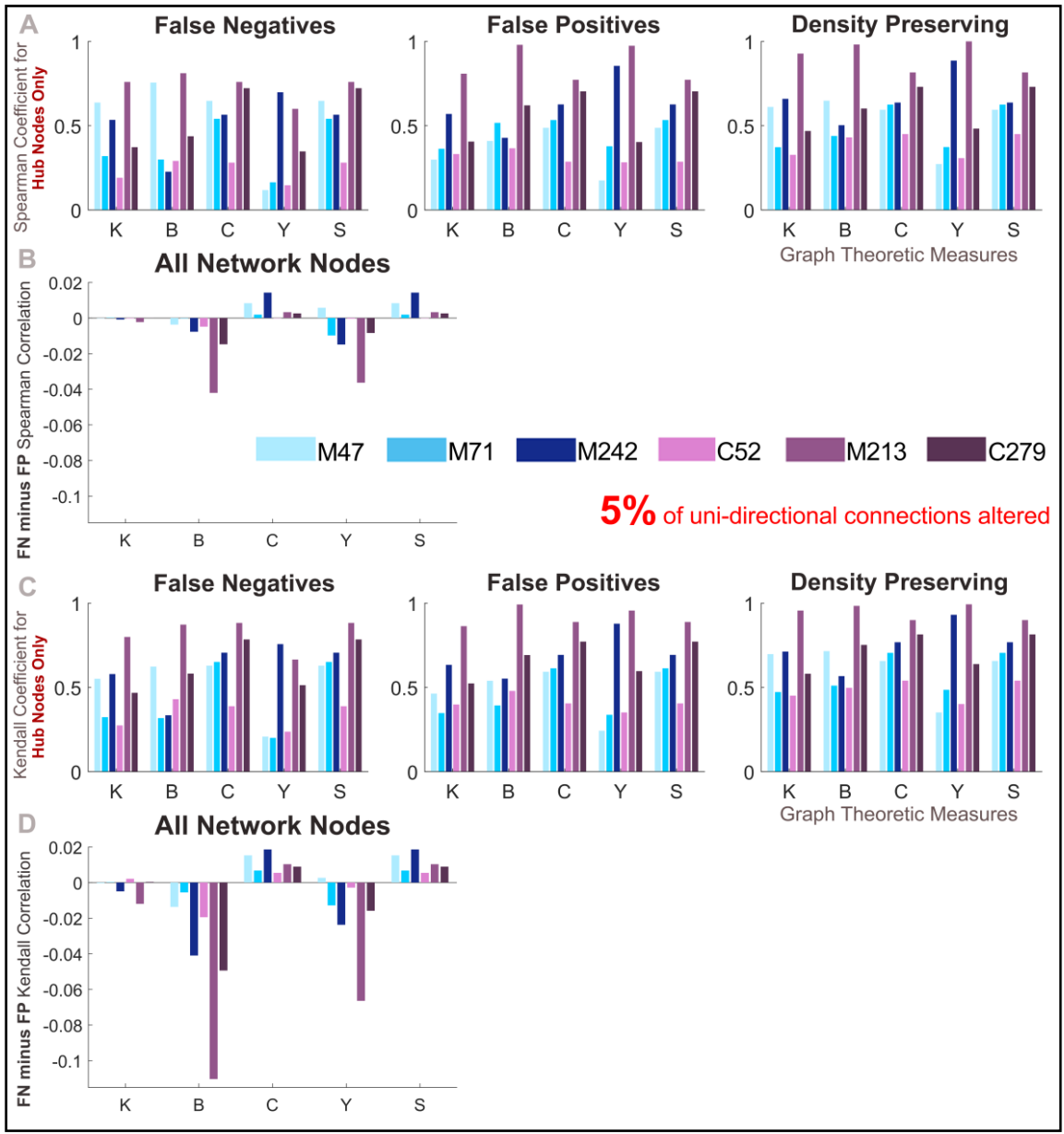

Supplementary Figure 7: **Nodal changes measured by the Spearman correlation and Kendall coefficient.** (A) Spearman correlation of hub nodes across all perturbed networks, for each graph-theoretic measure. (B) Difference in the spearman correlation between the false-negative and false-positive networks for all nodes in the network. A positive value indicates the false-negative connections cause greater changes in the ranking of nodes, whereas a negative value indicates the same for false-positive connections. (C) Kendall coefficient of hub nodes across all perturbed networks, for each graph-theoretic measure. (D) Difference in the Kendall coefficient between the false-negative and false-positive networks for all nodes in the network. (A-D) Results correspond to the mean over 50 trials for which 5% of randomly selected uni-directional connections are modified in each perturbed network. Graph-theoretic measures are as follows: K=Degree, B=Betweenness centrality, C=Clustering coefficient, Y=Participation index and S=Small-world index ($S_i^{\rightarrow}$). M47: the macaque connectome with 47 nodes, M71: macaque N=71, M242: macaque N=242, C52: cat, M213: mouse, C279: C. elegans.

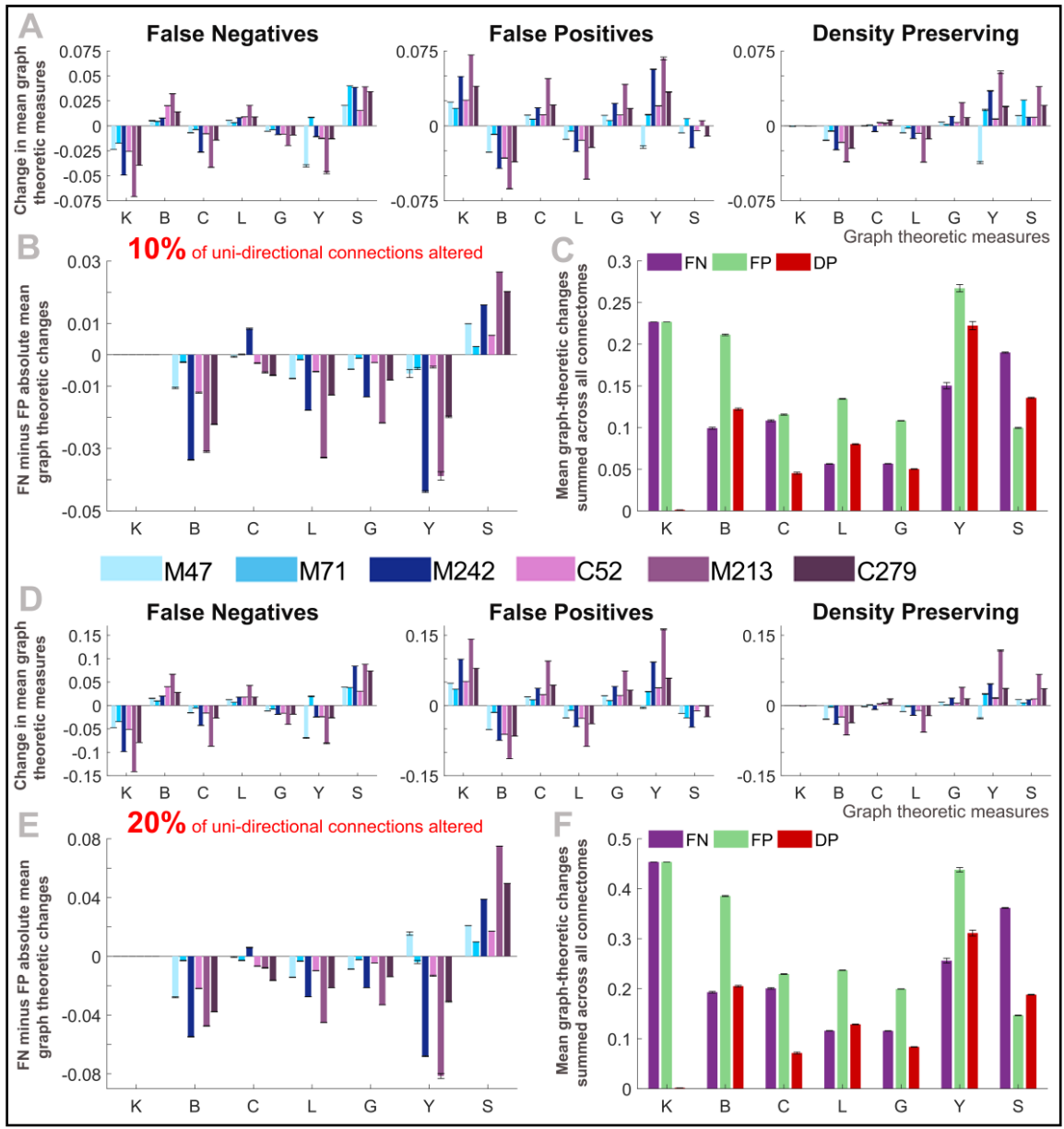

Supplementary Figure 8: **Relative change in mean graph-theoretic measures from the empirical connectomes for perturbed networks with 10% and 20% of uni-directional connections altered.** Changes in mean graph-theoretic measures across all connectomes and each type of perturbed network with 10% of uni-directional connections altered (A) and 20% of uni-directional connections altered (D). Difference between the changes in mean graph-theoretic measures for the 10% (B) and 20% (E) false-negative and false-positive networks. Mean changes in graph-theoretic measures for each of the perturbed networks with 10% of connections altered (C) or 20% of connections altered (F) summed across all connectomes. (A-F) The results represent the mean of these networks over 50 trials, and describe the change in the mean graph-theoretic measure (from the empirical to perturbed network) normalized by the mean of the empirical network (error bars show the standard error of the mean). Graph-theoretic measures are as follows: K=Degree, B=Betweenness centrality, C=Clustering coefficient, L=Characteristic path length, G=Global efficiency, Y=Participation index and S=Small-world index ($S^{\rightarrow}$, this measure is the mean over 1000 trials). M47: the macaque connectome with 47 nodes, M71: macaque N=71, M242: macaque N=242, C52: cat, M213: mouse, C279: C. elegans.

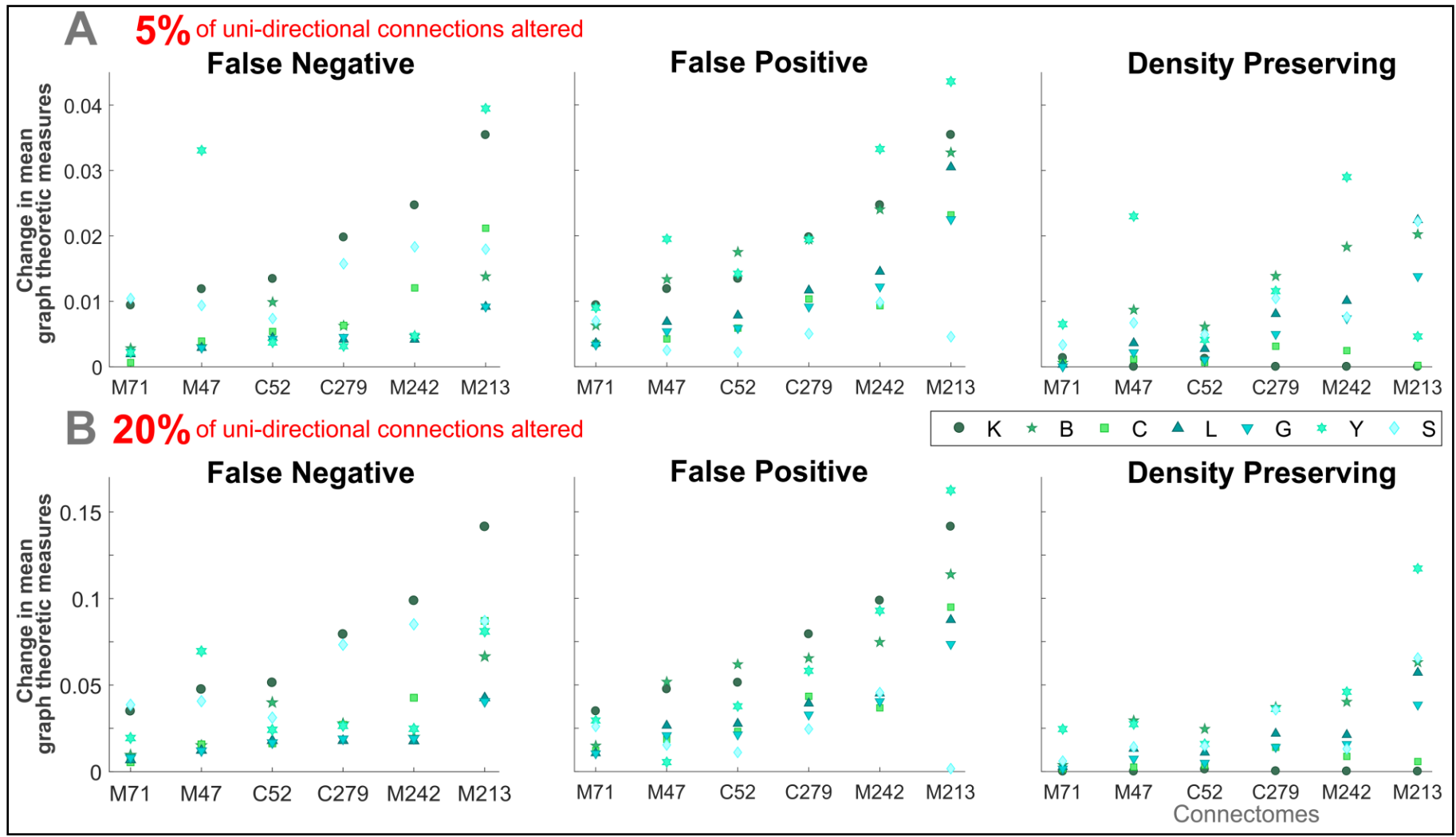

Supplementary Figure 9: **The change in mean graph-theoretic measures across connectomes sorted by the proportion of uni-directional connections (M71 = 17%, M47 = 24%, C52 = 26%, C279 = 40%, M242 = 49%, M213 = 71%).** (A) Relative change in mean graph theoretic measures (see legend) for each type of perturbed network with 5% of uni-directional connections altered. (B) Results for perturbed networks with 20% of uni-directional connections altered. The results represent the mean of these networks over 50 trials, and describe the change in the mean graph-theoretic measure (from the empirical to perturbed network) normalized by the mean of the empirical network. The legend shows each graph theoretic measure as follows: K=Degree, B=Betweenness centrality, C=Clustering coefficient, L=Characteristic path length, G=Global efficiency, Y=Participation index and S=Small-world index ($S^{\rightarrow}$, this measure is the mean over 1000 trials). M47: the macaque connectome with 47 nodes, M71: macaque N=71, M242: macaque N=242, C52: cat, M213: mouse, C279: C. elegans
.

| Connectome | | Macaque | Macaque | Macaque | Cat | Mouse | C. elegans |
|---|---|---:|---:|---:|---:|---:|---:|
| **Number of** | Nodes | 47 | 71 | 242 | 52 | 213 | 279 |
| | Modules | 4 | 4 | 5 | 3 | 5 | 4 |
| | Connections | 505 | 746 | 4090 | 818 | 3301 | 4903 |
| **Proportion of** | Uni-directional connections | 0.24 | 0.17 | 0.49 | 0.26 | 0.71 | 0.40 |
| **Density of** | All connections | 0.23 | 0.15 | 0.07 | 0.31 | 0.07 | 0.06 |
| | Uni-directional connections | 0.06 | 0.03 | 0.03 | 0.08 | 0.05 | 0.03 |
| **Number of connections** | Inter-/intra-modular ratio | 0.51 | 0.39 | 0.63 | 0.53 | 0.58 | 0.45 |
| | Core | 263 | 337 | 2080 | 362 | 1025 | 1763 |
| | Feeder | 194 | 339 | 1705 | 346 | 1729 | 2117 |
| | Peripheral | 48 | 70 | 305 | 110 | 547 | 1023 |
| **Proportion of uni-directional connections** | Core | 0.19 | 0.12 | 0.43 | 0.15 | 0.63 | 0.37 |
| | Feeder | 0.29 | 0.21 | 0.56 | 0.39 | 0.76 | 0.43 |
| | Peripheral | 0.29 | 0.29 | 0.59 | 0.20 | 0.70 | 0.37 |
| **Density of** | Hub-hub connections | 0.69 | 0.57 | 0.43 | 0.85 | 0.27 | 0.35 |
| | Hub-peripheral connections | 1.64 | 1.55 | 0.75 | 1.62 | 1.37 | 0.66 |
| **Proportion of uni-directional connections** | Hub-hub connections | 0.16 | 0.05 | 0.30 | 0.05 | 0.54 | 0.31 |
| | Hub-peripheral connections | 0.20 | 0.19 | 0.43 | 0.26 | 0.84 | 0.36 |
| **Mean** | Degree | 21.5 | 21.0 | 33.8 | 31.5 | 31.0 | 35.1 |
| | Betweenness centrality | 48 | 92 | 368 | 41 | 335 | 417 |
| | Clustering coefficient | 0.58 | 0.47 | 0.37 | 0.59 | 0.28 | 0.29 |
| | Participation index | 0.378 | 0.288 | 0.380 | 0.381 | 0.325 | 0.332 |
| | Small-world index | 2.23 | 2.78 | 4.71 | 1.79 | 3.31 | 4.20 |
| **Network measures** | Characteristic path length | 2.05 | 2.33 | 2.53 | 1.81 | 2.63 | 2.50 |
| | Global efficiency | 0.57 | 0.50 | 0.44 | 0.64 | 0.43 | 0.44 |

Supplementary Table 1: **Network characteristics of each empirical connectome.** Details of modularity determination are presented in Methods.

|  | Proportion of randomly selected connections | Modularity in perturbed networks | Number of trials | Connectomes | Definition of hubs in perturbed networks |
| --- | --- | --- | --- | --- | --- |
| Fig. 1 | - | - | - | All | - |
| Fig. 2 | 100% | Redefined | 1 | M47 | - |
| Fig. 3 | 100% | Redefined | **A**: 1, **B**: 1 | All | - |
| Fig. 4 | 100% | Redefined | **A**, **C**&**D**: 1, **B**&**E**: 1000 | **A**, **C**&**D**: M47, **B**&**E**: All | **A**&**C**-**E**: Empirical, **B**: Redefined |
| Fig. 5 | 5% | Empirical | 50 | All | Empirical |
| Fig. 6 | 5% | Empirical & Redefined | 50, except *small world index*: 1000 | All | - |
| S.Fig. 1 | 0-100% | - | 10, 000 | All | - |
| S.Fig. 2 | 100% | - | 1 | **A**, **B**: M47, **C**: All | **A**, **B**: As per *in-/ out- degree* of Empirical network, **C**: Empirical |
| S.Fig. 3 | 100% | - | 1000 | All | Redefined |
| S.Fig. 4 | 100% | - | 50 | M47 | Empirical |
| S.Fig. 5 | 100% | Redefined | Plots: 1, Bars: 1000 | All | Empirical |
| S.Fig. 6 | 100% | Both Displayed | 50 | All | Redefined |
| S.Fig. 7 | 5% | Empirical | 50 | All | Empirical |
| S.Fig. 8 | 10% and 20% | Redefined | 50, except *small world index*: 1000 | All | - |
| S.Fig. 9 | 5% and 20% | Redefined | 50, except *small world index*: 1000 | All | - |

Supplementary Table 2: **Methodological details for analyses presented in each figure and in the supplementary material.** M47: the macaque connectome with 47 nodes.

| Graph-theoretic Measure | | Formula |
|---|---|---|
| **Degree** (Rubinov and Sporns, 2010) | $K_i$= degree of node *i* (sum of directed in- and out-degree) | (both) $K_i = \sum_{j \in N} a_{ij}$ <br> $K_i^{in} = \sum_{j \in N} a_{ji}$ $K_i^{out} = \sum_{j \in N} a_{ij}$ |
| **Betweenness Centrality** (Freeman, 1978) | $B_i$ = betweenness centrality of node *i*, $B_{hj}(i)$ = number of shortest paths between *h* & *j* passing through *i*, $B_{hj}$ = number of shortest paths between *h* & *j* | $B_i = \dfrac{1}{(n-1)(n-2)} \sum_{\substack{h,j \in N \\ h \neq j, h \neq i, j \neq i}} \dfrac{B_{hj}(i)}{B_{hj}}$ |
| **Number of Triangles** (Rubinov and Sporns, 2010) | $t_i^{\rightarrow}$ = number of triangles around node *i* | $t_i^{\rightarrow} = \dfrac{1}{2} \sum_{j,h \in N} (a_{ij} + a_{ji})(a_{ih} + a_{hi})(a_{jh} + a_{hj})$ |
| **Clustering Coefficient** (Fagiolo, 2007) | $C_i^{\rightarrow}$= clustering coefficient of node *i* <br><br> $C^{\rightarrow}$= mean clustering coefficient of the network | $C_i^{\rightarrow} = \dfrac{1}{n} \sum_{i \in N} \dfrac{t_i^{\rightarrow}}{(K_i^{out} + K_i^{in})(K_i^{out} + K_i^{in} - 1) - 2\sum_{i \in N} a_{ij} a_{ji}}$ <br><br> $C^{\rightarrow} = \sum_{i=1}^{N} \dfrac{C_i^{\rightarrow}}{N}$ |
| **Shortest Path Length** (Rubinov and Sporns, 2010) | $d_{ij}^{\rightarrow}$ = shortest path length between nodes *i* & *j*, where $g_{i \rightarrow j}$ is the shortest path between *i* & *j* | $d_{ij}^{\rightarrow} = \sum_{a_{ij} \in g_{i \rightarrow j}} a_{ij}$ |
| **Characteristic Path Length** (Watts and Strogatz, 1998) | $L^{\rightarrow}$ = average distance between all nodes | $L^{\rightarrow} = \dfrac{1}{n} \sum_{i \in N} \dfrac{\sum_{j \in N, j \neq i} d_{ij}^{\rightarrow}}{n-1}$ |
| **Global Efficiency** (Latora and Marchiori, 2001) | $G^{\rightarrow}$ = global efficiency of the network | $G^{\rightarrow} = \dfrac{1}{n} \sum_{i \in N} \dfrac{\sum_{j \in N, j \neq i} (d_{ij}^{\rightarrow})^{-1}}{n-1}$ |
| **Participation Index** (Guimera and Amaral, 2005) | $Y_i^{out}$ =out-participation index, $M$= set of modules, $K_i^{out}(m)$ = number of out- connections between *i* & all nodes in module *m* | $Y_i^{out} = 1 - \sum_{m \in M} \left( \dfrac{K_i^{out}(m)}{K_i^{out}} \right)^2$ |
| **Small Worldness** (Humphries and Gurney, 2008) | $S_i^{\rightarrow}$ = small worldness of node *i* <br> $S^{\rightarrow}$ = small world index of network <br> $C_{i\,rand}^{\rightarrow}$ = clustering of a random network, <br> $L_{rand}^{\rightarrow}$ = path length of a random network | $S_i^{\rightarrow} = \dfrac{C_i^{\rightarrow}/C_{i\,rand}^{\rightarrow}}{L^{\rightarrow}/L_{rand}^{\rightarrow}}$ <br><br> $S^{\rightarrow} = \dfrac{C^{\rightarrow}/C_{rand}^{\rightarrow}}{L^{\rightarrow}/L_{rand}^{\rightarrow}}$ |

**Notations:** $N$ = all nodes in the network, $n$ = number of nodes, $L$ = all connections, $l$ = number of connections, $(i, j)$ = connection between nodes *i* & *j*, ($i, j \in N$), $a_{ij}$ = connection status between *i* & *j*: = 1 when a connection from node *i* to *j* exists, otherwise = 0, $l = \sum_{i,j \in N} a_{ij}$ (each bidirectional connection is counted twice, as $a_{ij}$ & as $a_{ji}$), *rand* = random network, $\rightarrow$ indicates formulae that consider directionality

Supplementary Table 3: **Reference, description and formula for each graph-theoretic measure used in this study.** Table adapted from Rubinov and Sporns (2010).

|  | **M47** | **M71** |  | **M242** |  |  | **C52** |  | **M213** |  |  | **C279** |  |  |  |
|---|---|---|---|---|---|---|---|---|---|---|---|---|---|---|---|
| **Connector** | V4 | A46 | 46 | 7b | 23 | 25 | CGp EPp | GPi | LGv | MH | AVAR | PVPL | AIBL | RMGL |
|  | FEF | TF | 24 | 8A | S2 | 46v | 35 | STN | VM | MPT | AVAL | AVDL | RIAL | AVG |
|  | 46 | TPT | LIP | 23c | 24c | 24b | AES | CLA | RM | PA | AVBR | PVT | RIBL | DVC |
|  | 7a | A7a | 32 | PIT | PS |  | 36 | CLI | MM | PTLp | AVBL | AVHR | RIBR | HSNR |
|  | TF | V4 | TF | 10 | 6M |  | Ia | PP | NOT | SPFm | DVA | PVR | AVJR | RIMR |
|  | 5 | FEF | 13a | TE | ENT |  | Ig | PPN | MGd | BLA | PVCR | PVNR | RMDL | RIH |
|  | FST | TS3 | 12o | PGm | 8B |  | 7 | PERI | RCH | VISam | PVCL | AIBR | AVKL | AVL |
|  | MT | ER | 9 | 13 | F5 |  | 6m | PT | ILA |  | AVER | PVPR | AVJL |  |
|  |  | TH | TH | 36 | 14 |  | 5Al | ACAd | SUBd |  | AVDR | AVHL | RIML |  |
|  |  |  | 11 | F7 | Iai |  | 20a | SPFp | LGd |  | AVEL | RIAR | RIGL |  |
| **Provincial** | 7b | A7b LIP | 12l |  |  |  |  | LGd | PAA |  | RIML | RIH |  |  |

Supplementary Table 4: **Connector and provincial hubs identified for each connectome.** The hub definition based on the degree is K greater than one standard deviation above the mean and further classification of hubs based on the participation index for connector hubs is Y > 0.35 and provincial hubs is Y ≤ 0.35. Each section (and column) is sorted by the highest degree. M47: the macaque connectome with 47 nodes, M71: macaque N=71, M242: macaque N=242, C52: cat, M213: mouse, C279: C. elegans.